\documentclass[12pt]{article}
\usepackage{epsfig,amssymb,amsfonts}
\usepackage{amsthm}
\date{\today}
\begin{document}
\title{\bf Diffusion of Power in Randomly Perturbed Hamiltonian
 Partial Differential Equations}
\author{
  E. Kirr \thanks{
 Department of Mathematics, University of Chicago}
 \hspace{.05 in}
 and
 M.I. Weinstein \thanks{
 Mathematical Sciences Research, Bell Laboratories - Lucent Technologies,
  Murray Hill, NJ}
}
   \baselineskip=18pt
   \maketitle
\def\diag{\mathop{\rm diag}}
\def\M{(\varepsilon^2\Gamma)^{-\alpha}}
\def\MM{(\varepsilon^2\Gamma/2)^{-\alpha}}
\def\un{\underline}
\def\nn{\nonumber}
\newcommand{\no}{\nonumber}
\newcommand{\be}{\begin{equation}}
\newcommand{\ee}{\end{equation}}
\newcommand{\ba}{\begin{eqnarray}}
\newcommand{\ea}{\end{eqnarray}}
\newcommand{\ve}{\varepsilon}
\newcommand{\nit}\noindent
\newcommand{\D}\partial
\def\Pc{{\bf P_c}}
\def\tA{{\tilde A}}
\newcommand{\lan}\langle
\newcommand{\ran}\rangle
\newcommand{\wpl}{w_+}
\newcommand{\wmi}{w_-}
\newcommand{\bXYZ}{{\bf XYZ}}
\newcommand{\spectrum}{\sigma_{\rm cont}(H_0)}
\newcommand{\cO}{{\cal O}}
\newcommand{\dm}{dm}
\newcommand{\cT}{{\cal T}}
\newcommand{\cA}{{\cal A}}
\newcommand{\la}{\lambda}
\newcommand{\ra}{\rightarrow}
\newcommand{\Z}{\mathbb{Z}}
\newcommand{\N}{\mathbb{N}}
\newcommand{\res}{I_{res}}
\newcommand{\bP}{\mathbb{P}}
\newcommand{\E}{\mathbb{E}}
\newcommand{\Q}{\mathbb{Q}}
\newcommand{\R}{\mathbb{R}}
\newcommand{\h}{\mathbb{H}}
\newcommand{\C}{\mathbb{C}}
\newcommand{\T}{\mathbb{T}}
\newcommand{\I}{\mathbb{I}}
\newtheorem{theo}{Theorem}[section]
\newtheorem{defin}{Definition}[section]
\newtheorem{prop}{Proposition}[section]
\newtheorem{lem}{Lemma}[section]
\newtheorem{cor}{Corollary}[section]
\newtheorem{rmk}{Remark}[section]
\newtheorem{conjecture}[theo]{Conjecture}
\newcommand{\pulse}{%
  \begin{picture}(60,50)(0,-20)
  \qbezier(0,0)(5,0)(10,15)
  \qbezier(10,15)(15,30)(20,30)
  \qbezier(20,30)(22,30)(25,18)
  \qbezier(25,18)(27,10)(30,10)
  \qbezier(30,10)(34,10)(38,15)
  \qbezier(38,15)(42,20)(45,20)
  \qbezier(45,20)(49,20)(53,10)
  \qbezier(53,10)(57,0)(60,0)
  \multiput(0,0)(0,-2){10}{\line(0,-1){1}}
  \multiput(60,0)(0,-2){10}{\line(0,-1){1}}
  \put(30,-20){\vector(-1,0){30}}
  \put(30,-20){\vector(1,0){30}}
  \put(0,-15){\makebox(60,10){{\tiny $T$}}}
  \end{picture}}

\begin{abstract}
{\bf Abstract}
We study the evolution of the energy (mode-power) distribution
 for a class of randomly perturbed
 Hamiltonian partial differential equations
  and derive {\it master equations} for the dynamics
of the expected power in the discrete modes.
 In the case where the unperturbed dynamics has only
 discrete frequencies (finitely or infinitely many) the mode-power distribution
 is  governed
by an equation of discrete diffusion type for times of order $\cO(\ve^{-2})$.
Here $\ve$ denotes the size of the random perturbation.
 If the unperturbed system has
discrete and continuous spectrum the mode-power distribution is governed
by an equation of discrete diffusion-damping type for times of order
 $\cO(\ve^{-2})$. The methods involve an extension of the authors' work
on deterministic periodic and almost periodic perturbations, and yield new
 results which complement results of others, derived  by probabilistic methods.
\end{abstract}
\tableofcontents
\section{Introduction}\label{ch:diff}

The evolution of an arbitrary initial condition of
 linear autonomous Hamiltonian partial differential equation
(Schr\"odinger equation),
\be i\D_t\phi=H_0\phi,\label{eq:unperturbed}\ee
where $H_0$ is self-adjoint operator,
can be studied by decomposing the initial
state in terms of the eigenstates (bound modes) and
generalized eigenstates (radiation or continuum modes) of $H_0$.
The mode amplitudes evolve independently according to a system of decoupled
ordinary differential equations and the energy or power in each mode, the
square of the mode amplitude, is independent of time.
  If the system (\ref{eq:unperturbed})
is perturbed
\be i\D_t\phi=(H_0+W(t))\phi,\label{eq:perturbed}\ee
where $W(t)$  respects the Hamiltonian
 structure ($W^*=W$),  then
the system of ordinary differential equations typically becomes
an infinite coupled system of equations, so-called coupled mode equations.
If $W(t)$ has general time-dependence (periodic, almost periodic, random,...),
 the  solutions of the coupled mode equations  can exhibit
very complex behavior. Of fundamental importance is the question how
the mode-powers evolve with $t$.
Kinetic equations, which govern their evolution are called
{\it master equations}
\cite{kn:VanHove}, \cite{davies76:opensystems} and go back to the work of Pauli
\cite{kn:Pauli-1928}. A general approach to stochastic systems
is presented in \cite{kn:papa1,kn:papa2,kn:papa3,kn:kp}; see also
\cite{leb:diff,jg:ahar,jg:light}.
 Master equations have been derived in many contexts
in statistical mechanics, ocean acoustics and
  optical wave-propagation in waveguides.

We present a theory of power evolution for
 (\ref{eq:perturbed}), for a class of perturbations, $W(t)$,
which are random in $t$. Our theory handles the case where $H_0$ has spectrum
consisting of bound states (finitely or infinitely many discrete eigenvalues)
 and radiation modes (continuous spectrum). It is
a natural extension of the
analysis in our work on deterministic periodic, almost periodic
and nonlinear systems; see, for example,
 \cite{kn:mwk,kw:mbs,kw:apt,kn:rdamping}.
Our approach is complementary
to the probablistic approach of \cite{jg:ahar,jg:light,kn:papa2,kn:papa3,kn:kp}. The
model we consider is
well-suited to the study of the effects of an ``engineered'' perturbation
of the system, {\it e.g.} a prescribed train of light pulses incident on
an atomic system, or prescribed distribution of defects encountered by
waves propagating along a waveguide; see below.
 We also give very
 detailed information on the energy transfer between the subsystems governed
by discrete ``oscillators'' and  continuum ``radiation field''.

In particular, we study the problem
 \be\label{eq:3pert} i\D_t\phi\
=\ \left(\ H_0\ +\ \ve g(t)\beta\ \right)\phi,\quad
 \ee
where $\ve$ is small, and  $H_0$ and $\beta$ are self-adjoint operator on
the Hilbert space ${\cal H}$. $H_0$ is assumed to support finitely
or infinitely many bound states.
 For example, $H_0=-\Delta+V(x)$, where $V$ is smooth
 and sufficiently rapidly decaying as $|x|\to\infty$.
 $\beta $ is assumed to be bounded. $g(t)$ is a real
valued function of the form of a sequence of short-lived perturbations
  or ``defects''; see figure 1.
Our methods can treat the case of more general perturbations, {\it e.g.}
  $W(t,x)=\beta(t,x)$, but to simplify the presentation we consider
the separable case $W(t)=g(t)\beta(x)$.

Models of the above type arise natural in many contexts. Among them are
the interaction between an atom and a train of light pulses
 \cite[and references therein]{ryb:des} , a field of great
current interest in the control of quantum systems.
Such trains of localized  perturbations also  model
sequences of localized defects along waveguides, see \cite{kn:Marcuse},
\cite{kn:MSW}, introduced by accident or design.

We construct $g(t)$ as follows. Start with $g_0(t)$, a
 fixed real-valued function with support contained in the interval $[0,T]$
 and let  $\{d_j\}_{j\ge0}$
 be a nonnegative sequence. Define
\be
g(t)\ =\ \sum_{n=0}^\infty\ g_0(t-t_n),\ {\rm where}\label{eq:gdef}
\ee
\ba
t_0&=&d_0\nn\\
t_n &=& (d_0+T)+(d_1+T)+\dots+(d_{n-1}+T)+d_n, n\ge 1
\label{eq:tndef} \ea denotes the onset of the $n^{th}$ defect.
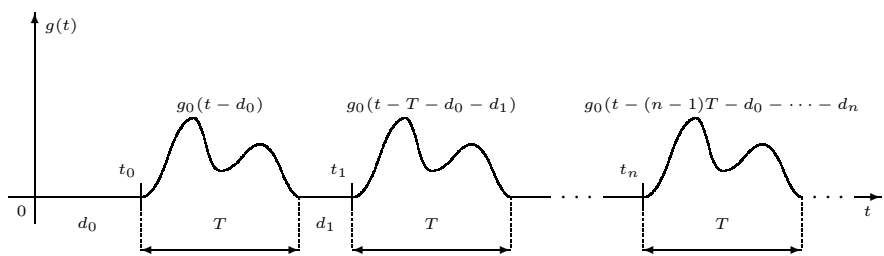
\begin{figure}[h]
 \begin{center}
  \begin{picture}(330,90)(-10,0) \setlength{\unitlength}{1pt}
    \put(0,20){\line(1,0){40}}
    \put(40,15){\line(0,1){10}}
    \put(30,25){\makebox(10,10){{\tiny$t_0$}}}
       \put(0,5){\makebox(40,10){{\tiny $d_0$}}}
    \put(40,0){\pulse}
       \put(40,50){\makebox(60,10){{\tiny $g_0(t-d_0)$}}}
    \put(100,20){\line(1,0){20}}
    \put(120,15){\line(0,1){10}}
    \put(110,25){\makebox(10,10){{\tiny$t_1$}}}
     \put(100,5){\makebox(20,10){{\tiny $d_1$}}}
    \put(120,0){\pulse}
       \put(120,50){\makebox(60,10){{\tiny $g_0(t-T-d_0-d_1)$}}}
    \put(180,20){\line(1,0){15}}
    \multiput(200,20)(5,0){3}{\circle*{1}}
    \put(215,20){\line(1,0){15}}
    \put(230,15){\line(0,1){10}}
    \put(220,25){\makebox(10,10){{\tiny$t_n$}}}
    \put(230,0){\pulse}
       \put(230,50){\makebox(60,10){{\tiny $g_0(t-(n-1)T-d_0-\cdots -d_n$}}}
    \multiput(295,20)(5,0){3}{\circle*{1}}
    \put(310,20){\vector(1,0){10}}
      \put(310,10){\makebox(10,10){{\tiny $t$}}}
    \put(0,10){\vector(0,1){80}}
    \put(0,20){\line(-1,0){10}}
    \put(-10,10){\makebox(10,10){{\tiny $0$}}}
    \put(0,80){\makebox(20,10){{\tiny $g(t)$}}}
\end{picture}
\end{center}
\caption{Train of short lived perturbations or ``defects''.
 The onset time for the $n^{th}$ defect,
 $t_n$, is given by (\ref{eq:tndef}).}
\label{fig:train}
\end{figure}


Note that, if the sequence $\{d_j\}_{j\ge0}$ is periodic 
then $g(t)$ is periodic. 
In this case, the system
(\ref{eq:3pert}) has already been analyzed by
 time-independent methods \cite{kn:YA} or,
more recently and under less restrictive hypothesis, in \cite{kn:mwk,kw:mbs}.
For $\{d_j\}_{j\ge0} $ quasiperiodic or almost periodic
 (see \cite{corduneanu,simon:schroedinger}
for a definition) the situation is more delicate.
In \cite{kw:mbs} we treat a general class of almost
 periodic perturbations of the form:
\be W(t)\ =\ \sum_{j\ge0}\cos(\mu_j t)\beta_j,\label{eq:W-kw-mbs}
\ee with appropriate ``small denominator'' hypotheses on the
frequencies $\{\mu_j\}$. We leave it for a future paper
\cite{kw:apt} to consider the case of almost periodic
$\{d_j\}_{j\ge0} $ and to explore the connection with the results
in \cite{kw:mbs}. We note that a particular case has already been
treated in \cite[Appendix E]{kirr:thesis}.

Note that in  \cite{leb:diff} and \cite{kn:papa1}
 the numbers  $d_0,d_1,\ldots ,$ are equal to a fixed
constant and $g_0(t)$ is random while in our model $d_0,d_1,\ldots
,$ are random and $g_0(t)$ is fixed. This is another sense, in which
 our results complement those in
 the existing literature.

%
%

The paper is divided in two parts. The first part treats
stochastic perturbations of Hamiltonian systems with discrete
frequencies and then second part extends these results to the case
where the unperturbed system has discrete and continuous
frequencies. The stochastic perturbation is of order $\ve$ and
then the vector $P(\tau)\in \ell^1$, whose components are the
expected values of the squared discrete mode amplitudes
(mode-powers), satisfies on time scales $t=\cO(\ve^{-2})$ or
equivalently $\tau=\cO(1)$, the master equations of diffusion or
diffusion-damping type. Specifically, if $H_0$ has only discrete
spectrum (finite or infinite) then \be \D_\tau P(\tau) =
-BP(\tau),\ \ B\ge0 \nn \ee which has the character of a discrete
diffusion equation, {\it i.e.} \be \sum_k P_k(\tau)\ =\ \sum_k
P_k(0),\ \ \frac{d}{d\tau} P\cdot P =
 -\lan P,BP\ran\ \le0.
\label{eq:heat1}
\ee
If $H_0$ has both discrete and continuous spectra, then
\be
\D_\tau P(\tau) = (-B-\Gamma)P(\tau),\ \ B\ge0,\ \Gamma=\diag(\gamma_k)>0
\label{eq:heat2}
\ee
for which
\be \sum_k P_k(\tau)\ \le\ e^{-\gamma\tau}\sum_k P_k(0),
\nn\ee
where $\gamma=\min_k\ \gamma_k $.
\bigskip

In sections 2 and
3 we study (\ref{eq:3pert}) under the hypothesis that $H_0$
has no continuous spectrum (i.e. no radiation modes) and in section
4 we generalize to the case where $H_0$ has discrete and continuous spectrum.
  In section
\ref{se:norad} we present the main hypotheses on $H_0$ and
$g_0(t)$ and study the effect of a single short lived
perturbation. In section \ref{se:ex} we present our hypotheses on
$d_0,d_1,\ldots ,$ and analyze the effect of a train of
perturbations (\ref{eq:3pert}-\ref{eq:gdef}). We show that if
$d_0,d_1,\ldots ,$ are independent random variables with certain
distributions, see Hypothesis {\bf (H4)} and Examples 1 and 2,
diffusion occurs in the expected value for the powers of the
modes. Specifically, if we start with energy in one mode, then, on
a time scale of order $1/\ve^2$, one can expect the energy to be
distributed among all the modes. In section \ref{se:yesrad} we
analyze equation (\ref{eq:3pert}) under the hypothesis that $H_0$
has both discrete and continuous spectrum (i.e. supports both
bound modes and radiation modes). We prove a result similar to the
nonradiative case but now bound state- wave resonances lead to
loss of power. The effect of our randomly distributed
deterministic perturbation is very similar to the one induced by
purely stochastic perturbations, see
\cite{leb:diff,kn:kp,kn:papa2}, but quite different from the
effects of time almost periodic perturbations, see \cite{kn:mwk,kw:mbs}.
Section \ref{se:comp} is dedicated to such comparisons.

\nit {\bf Notation}

\nit 1) $\lan x \ran\ =\ \sqrt{1+x^2}$

\nit
 2) Fourier Transform:
\be \hat g(\xi)\ = \int_{-\infty}^\infty e^{-i\xi t}g(t)\ dt
\label{eq:FT}
\ee

\nit 3) We write $\zeta + c.c.$ to mean $\zeta+\bar\zeta$, where $\bar\zeta$
denotes the complex conjugate of $\zeta$.

\nit 4) $w'$ denotes the transpose of $w$.

\nit 5) $\lfloor q\rfloor$
 denotes the integer part of $q$.
\medskip

\nit{\bf Acknowledgement:} We would like to thank 
 G. C. Papanicolaou and J. L. Lebowitz  
for helpful discussions concerning this work. E.K. was supported in part
by the ASCI Flash Center at the University of Chicago.


\section{Short lived perturbation of a
system with discrete frequencies}\label{se:norad}

In this section we consider the
 perturbed dynamical system
\be\label{eq:schrodinger}
i\D_t\phi(t)=H_0\phi(t)+\ve g_0(t)\beta\phi(t,x),
\ee
where $H_0$ has only discrete spectrum and $g_0(t)$ is a short-lived (compactly
supported) function. We study the effect of this
 perturbation on the distribution
of energy among the modes of $H_0$. Here and in section \ref{se:yesrad} we are
extending the results in \cite{kn:ionize} to multiple bound states but under an
additional ``incoherence''  assumption; see (\ref{eq:IC}).

{\bf Hypotheses on $H_0,\ \beta$ and $g_0(t)$}

{\bf (H1)} $H_0$ is a self adjoint operator on a Hilbert space ${\cal H}$. It
has a pure point spectrum formed by the eigenvalues : $\{\la_j\}_{j\ge1}$ with
a complete set of orthonormal eigenvectors: $\{\psi_j\}_{j\ge1}:$
\be H_0\psi_j=\lambda_j\psi_j\ ,\ \lan\psi_i,\psi_j\ran\ =\ \delta_{ij}
\label{eq:eig}
\ee

{\bf (H2)} $\beta$ is a bounded self adjoint operator on ${\cal H}$ and
satisfies $\|\beta\|=1$.

{\bf (H3)} $g_0(t)\in L^2(\R)$ is real valued, has compact support contained
in $[0,T]$ on the positive real line and its $L^1$-norm, denoted by $\|g_0\|_1$
is $1$. Thus its Fourier transform has $L^\infty$-norm bounded by $1$.

Note that one can always take $\|\beta\|=1$ and $\|g_0\|_1=1$ by
 setting $\ve = \|g_0\|_1\cdot \|\beta\|$,
thus incorporating the size of  $g_0\beta$ in $\ve$. Therefore,
under assumptions (H2-H3), $\ve$ in (\ref{eq:schrodinger})
measures the actual size of the perturbation in the $L^1(\R,{\cal
H})$ norm. Our results are perturbative in $\ve$ and are valid for
$\ve$ sufficiently small.

By the standard contraction method one can show that (\ref{eq:schrodinger}) has
an unique solution $\phi(t)\in {\cal H}$ for all $t\in\R .$ Moreover, because
both $H_0$ and $g_0(t)\beta$ are self adjoint operators, we have for all
$t\in\R :$
\be\label{eq:noradnormb}
\|\phi(t)\|=\|\phi(0)\|.\ee
We can write $\phi(t)$ as a sum of projections onto the complete set of
orthonormal eigenvectors of $H_0 :$
\be\label{eq:modedecomposition}
\phi(t,x)=\sum_{j}a_j(t)\psi_j(x),
\ee
By Parseval's relation
\be\label{eq:parseval}
\sum_j\left| a_j(t)\right|^2=\|\phi(t)\|^2\equiv\|\phi(0)\|^2\ee
Now (\ref{eq:schrodinger}) can be rewritten as
\be
i\D_ta_k(t)=\la_ka_k(t)+\ve g_0(t)\sum_{j}a_j(t)\lan\psi_k,\beta\psi_j\ran ,\ k\in
\{1,2\ldots \}\label{eq:amplitudesystem}
\ee
where $\lan\cdot,\cdot\ran$ denotes the scalar product in ${\cal H}$.

Hence the equation (\ref{eq:schrodinger}) is equivalent to a weakly coupled
linear system in the amplitudes: $a_1,a_2,\ldots ,$ (\ref{eq:amplitudesystem}).

%
%

Since the perturbation size is $\ve$ we expect, in general, that
the change in energy in the $k^{\rm th}$ mode,
$|a_k(t)|^2-|a_k(0)|^2,$ to be of order $\ve.$ However with a
suitable random initial condition we can prove more subtle
behavior.

Suppose that there exists an averaging procedure applicable to the
amplitudes: $a_1,a_2,\ldots $ of the solutions of (\ref{eq:schrodinger}),
denoted by
$$
a(t)\mapsto \E (a(t))\in\C.
$$

We now state a fundamental result, applied throughout this paper,
   for a single defect which is
  compactly supported in time:

\begin{theo}\label{th:onepert} Assume the conditions {\bf (H1)-(H3)} hold
and the initial values for (\ref{eq:schrodinger}) are such that
 \be\label{eq:IC}
 \E\left( a_j(0)\overline{a_k(0)}\right)=0\qquad {\rm whenever}\ j\neq k.
\ee Then for all $t>\sup\{s\in\R\ |\ g_0(s)\neq 0\}$ and
$k\in\{1,2,\ldots \}$ we have \be\label{eq:powersystem}
P_k(t)-P_k(0)=\ve^2\sum_{j}|\alpha_{kj}|^2|\hat
g_0(-\Delta_{kj})|^2 (P_j(0)-P_k(0))+\cO (\ve^3), \ee where
$$P_k(t)\equiv \E\left(|a_k(t)|^2\right)$$ denotes the average
power in the $k^{th}$-mode at time $t$, $\alpha_{kj}\equiv
\lan\psi_k,\beta\psi_j\ran,$ $\hat g_0$ denotes the Fourier
transform of $g_0$ and $\Delta_{kj}\equiv \la_k-\la_j$.
\end{theo}

Note that (\ref{eq:powersystem}) can be written in the form:
\be\label{eq:powersystem2} P_k(t)=T_{\ve}P_k(0)+\cO (\ve^3), \ee
where
\be T_\ve=\I-\ve^2B;\qquad B\ge 0\label{eq:nTdef}\ee
$\I$ is
the identity operator (matrix) and $B$ is given by \be
B=\left(b_{kj}\right)_{1\le k,j},\qquad
b_{kj}=\left\{\begin{array}{lr} -|\alpha_{kj}|^2|\hat
g_0(-\Delta_{kj})|^2,& {\rm for}\ j\neq
k,\\
\sum_{l,l\neq k}|\alpha_{kl}|^2|\hat g_0(-\Delta_{kl})|^2,& {\rm
for}\ j=k\end{array}\right.\label{eq:rBdef} \ee In section
\ref{se:ex} we will discuss and use the properties of $B$ and
$T_\ve .$

 \nit\un{Proof of Theorem \ref{th:onepert}.} In the amplitude system,
(\ref{eq:amplitudesystem}), we remove the fast oscillations by
letting \be\label{eq:rAdef} a_k(t)=e^{-i\la_k t}A_k(t), \ee Note
that by (\ref{eq:parseval}) \be\label{eq:Aparseval} \sum_j\left|
A_j(t)\right|^2\equiv\|\phi(0)\|^2\ee Now
(\ref{eq:amplitudesystem}) becomes \be\label{eq:Asystem}
i\D_tA_k(t)=\ve g_0(t)\sum_{j}\alpha_{kj}e^{i\Delta_{kj}t}A_j(t),
\ee where \ba
\Delta_{kj}&\equiv &\la_k-\la_j,\label{eq:Deltadef}\\
\alpha_{kj}&\equiv &\lan\psi_k,\beta\psi_j\ran=\overline \alpha_{jk}.
\label{eq:alphadef}
\ea
The above system leads to the following one in product of amplitudes,
$A_k(t)\overline A_l(t)$:
\ba
\D_t (A_k(t)\overline A_l(t))&=&i\ve g_0(t)\sum_{j}\alpha_{jl}e^{i\Delta_{jl}t}
A_k(t) \overline A_j(t)\no\\
&-&
i\ve g_0(t)\sum_{j}\alpha_{kj}e^{i\Delta_{kj}t}A_j(t)\overline A_l(t),\label{eq:mpsystem}
\ea
In the particular case $k=l$ we have the power equation for each mode:
\be
\D_t |A_k(t)|^2=i\ve g_0(t)\sum_{j}\alpha_{jk}e^{i\Delta_{jk}t}A_k(t)
\overline A_j(t)+c.c.\ .\label{eq:psystem}
\ee
Note that the sum in (\ref{eq:psystem}) commutes with time integral
and expected value operators. This is due to (\ref{eq:Aparseval}) and the
dominant convergence theorem, see for example \cite{dur:pro}. Indeed consider
$$f_m(t)=\sum_{j=1}^m\alpha_{jk}e^{i\Delta_{jk}t}
A_k(t) \overline A_j(t)g_0(t).$$ From (\ref{eq:modedecomposition}) we have for
all $t\in\R $
$$\lim_{m\rightarrow\infty}f_m(t)=\lan\phi(t),\beta\psi_k\ran a_k(t)g_0(t).$$
From (\ref{eq:Aparseval}) and the  Cauchy-Schwarz inequality
$|\lan a,b\ran|\le\|a\|\ \|b\|$, we have for all $t\in\R $
\be\label{eq:fmint}|f_m(t)|\le\|\phi(0)\|^2|g_0(t)|.\ee The right hand side of
(\ref{eq:fmint}) is integrable and the dominant convergence theorem
applies. A similar argument is valid for expected values. Therefore, from now
on, we are going to commute both time integrals and expected values
 with summations like the one in (\ref{eq:psystem}).

We integrate (\ref{eq:psystem}) from $0$ to
$t>\sup\{s\in\R\ |\ g_0(s)\neq 0\}$ and integrate by parts the right hand side. The result is:
\ba
|A_k(t)|^2-|A_k(0)|^2&=&i\ve\sum_{j}\alpha_{jk}\int_0^tg_0(s)e^{i\Delta_{jk}s}A_k(s)\overline A_j(s)+c.c.\no\\
&=&-i\ve \sum_{j}\alpha_{jk}\int_s^\infty g_0(\tau)e^{i\Delta_{jk}\tau }d\tau
A_k(s)\overline A_j(s)\left|_{s=0}^{s=t}\right.+c.c. \label{eq:1intbyparts}\\
&+&i\ve\sum_{j}\alpha_{jk}\int_0^t\int_s^\infty
g_0(\tau)e^{i\Delta_{jk}\tau }d\tau\D_s\left(A_k\overline
A_j\right)(s)ds+c.c.\ .\no \ea The boundary terms are \ba
\lefteqn{-i\ve \sum_{j}\alpha_{jk}\int_s^\infty
g_0(\tau)e^{i\Delta_{jk}\tau }d\tau A_k(s)\overline
A_j(s)\left|_{s=0}^{s=t}\right.+c.c.\ }\no\\&=&i\ve
\sum_{j}\alpha_{jk}\hat g_0(-\Delta_{jk})A_k(0)\overline
A_j(0)+c.c.,\label{eq:bt} \ea
where $\hat g_0$ denotes the Fourier
Transform of $g_0$; see (\ref{eq:FT}).
 Note that
upon taking the average, using (\ref{eq:IC}) and the fact that
$\hat g_0(0)$ is real, these boundary terms vanish.

Into the last term in (\ref{eq:1intbyparts}) we substitute
(\ref{eq:mpsystem}): \ba
\lefteqn{i\ve\sum_{j}\alpha_{jk}\int_0^t\int_s^\infty g_0(\tau)e^{i\Delta_{jk}\tau }d\tau \D_s\left(A_k\overline A_j\right)(s)ds=}\no\\
&=&+|\ve|^2\sum_{j,p}\alpha_{jk}\alpha_{kp}\int_0^t\int_s^\infty g_0(\tau)e^{i\Delta_{jk}\tau }d\tau
g_0(s)e^{i\Delta_{kp}s}A_p(s)\overline A_j(s)ds\no\\
&-&|\ve|^2\sum_{j,q}\alpha_{jk}\alpha_{qj}\int_0^t\int_s^\infty g_0(\tau)e^{i\Delta_{jk}\tau }d\tau
g_0(s)e^{i\Delta_{qj}s}A_k(s)\overline A_q(s)ds.\label{eq:e1}
\ea
We again integrate by parts both terms in (\ref{eq:e1}):
\ba
\lefteqn{i\ve\sum_{j}\alpha_{jk}\int_0^t\int_s^\infty g_0(\tau)e^{i\Delta_{jk}\tau }d\tau \D_s\left(A_k\overline A_j\right)(s)ds=}\no\\
&-&|\ve|^2\sum_{j,p}\alpha_{jk}\alpha_{kp}\int_u^\infty
g_0(s)e^{i\Delta_{kp}s}\int_s^\infty g_0(\tau)e^{i\Delta_{jk}\tau
}d\tau
dsA_p(u)\overline A_j(u)\left|_{u=0}^{u=t}\right.\no\\
&+&|\ve|^2\sum_{j,q}\alpha_{jk}\alpha_{qj}\int_u^\infty
g_0(s)e^{i\Delta_{qj}s} \int_s^\infty g_0(\tau)e^{i\Delta_{jk}\tau
}d\tau
dsA_k(u)\overline A_q(u)\left|_{u=0}^{u=t}\right.\no\\
&+&|\ve|^2\sum_{j,p}\alpha_{jk}\alpha_{kp}\int_0^t\int_u^\infty\int_s^\infty
g_0(\tau)e^{i\Delta_{jk}\tau}d\tau g_0(s)e^{i\Delta_{kp}s}ds\D_u\left(A_p\overline A_j\right)(u)du\no\\
&-&\!\!
|\ve|^2\sum_{j,q}\alpha_{jk}\alpha_{qj}\int_0^t\int_u^\infty\int_s^\infty
\!\! g_0(\tau)e^{i\Delta_{jk}\tau}d\tau
g_0(s)e^{i\Delta_{qj}s}ds\D_u\left(A_k\overline
A_q\right)(u)du\label{eq:2intbyparts}. \ea

Note that the boundary terms calculated at $``u=t"$ are zero since
$t>\sup\{s\in\R\ |\ g_0(s)\neq 0\}$. Upon taking the expected
value and using (\ref{eq:IC}) the only boundary terms contributing
are the ones for which $u=0$ and $j=p$ in the second row of
(\ref{eq:2intbyparts}):
\ba\lefteqn{\sum_{j}|\alpha_{kj}|^2\int_0^\infty\int_s^\infty
g_0(\tau)e^{i\Delta_{jk}\tau }d\tau
g_0(s)e^{i\Delta_{kj}s}ds\E\left(|A_j(0)|^2\right) +c.c.}\no\\
&=&\sum_{j}|\alpha_{kj}|^2\E\left(|A_j(0)|^2\right)\cdot
2\Re\int_0^\infty\int_s^\infty g_0(\tau)e^{i\Delta_{jk}\tau }d\tau
g_0(s)e^{i\Delta_{kj}s}ds\label{eq:sbterms}\ea and the ones for
which $u=0$ and $q=k$ in the third row of (\ref{eq:2intbyparts}):
\ba\lefteqn{\sum_{j}|\alpha_{kj}|^2\int_0^\infty\int_s^\infty
g_0(\tau)e^{i\Delta_{jk}\tau }d\tau
g_0(s)e^{i\Delta_{kj}s}ds\E\left(|A_k(0)|^2\right) +c.c.}\no\\
&=&\sum_{j}|\alpha_{kj}|^2\E\left(|A_k(0)|^2\right)\cdot
2\Re\int_0^\infty\int_s^\infty g_0(\tau)e^{i\Delta_{jk}\tau }d\tau
g_0(s)e^{i\Delta_{kj}s}ds\label{eq:tbterms}\ea To compute
(\ref{eq:sbterms}-\ref{eq:tbterms}) we use the lemma:

\begin{lem}\label{lem:rescalc} If $g_0(t), t\in\R$ is square
 integrable with compact support included in the positive
real line then for all $\lambda\in\R $ the following identity
holds
$$2\Re\int_0^\infty\int_s^\infty
g_0(\tau)e^{i\la\tau }d\tau g_0(s)e^{-i\la s}ds=|\hat g_0(-\la)|^2
.$$
\end{lem}

\nit\un{Proof.} For any $\la\in\R$ we have: \ba
\lefteqn{\int_0^\infty\int_s^\infty g_0(\tau)e^{i\la\tau }d\tau
g_0(s)e^{-i\la s}ds}\no\\
&=&\lim_{\ve\searrow 0}\int_0^\infty\int_s^\infty
g_0(\tau)e^{i(\la+i\ve )\tau }d\tau g_0(s)e^{-i\la s}ds
\no\\&=&{1\over 2\pi}\lim_{\ve\searrow 0}\int_0^\infty
g_0(s)e^{-i\la s}ds\int_{-\infty}^\infty\hat g_0(\mu)
\int_{s}^\infty e^{i(\la +\mu +i\ve)\tau}d\tau d\mu
\nn\\&=&{i\over 2\pi}\lim_{\ve\searrow 0}\int_0^\infty
g_0(s)e^{-i\la s}ds\int_{-\infty}^\infty\frac{\hat
g_0(\mu)}{\mu+\la+i\ve}e^{is(\mu+\la+i\ve)}d\mu \no\\
&=&{i\over 2\pi}\lim_{\ve\searrow 0} \int_{-\infty}^\infty\frac{\hat g_0(\mu)\hat g_0(-\mu)}{\mu+\la+i\ve}d\mu \no\\
&=&{1\over 2} |\hat g_0(-\la)|^2+{i\over 2\pi}{\rm
P.V.}\int_{-\infty}^\infty\frac{|\hat
g_0(\mu)|^2}{\mu+\la}d\mu\label{eq:2boundary}
 \ea The last relation in (\ref{eq:2boundary})
is the Plemelj-Sohotsky's formula for (temperate) distributions:
$$\lim_{\ve\searrow 0}\frac{1}{x+i\ve}=
{\rm P.V.}{1\over
x}-i\pi\delta(x)\stackrel{def}{=}\frac{1}{x+i0}.$$ Note that $\hat
g_0(\mu)\in C^\infty (\R )\cap L^2(\R).$ Since
(\ref{eq:2boundary}) is already decomposed in its real and
imaginary part the lemma follows. \ \ \ \ []

Into the triple integral terms of (\ref{eq:2intbyparts})
we again  substitute (\ref{eq:mpsystem}). Then one can show that the 1-norm of
this correction vector is dominated by $|\ve|^3\ \|g_0\|_1^3\
\|\beta\|^3\ \|\phi(0)\|^2.$ Hence, it is of order $\cO (|\ve
|^3)$.

Thus, after applying Lemma \ref{lem:rescalc}
  to (\ref{eq:sbterms}-\ref{eq:tbterms})
and using (\ref{eq:1intbyparts}) we arrive at the conclusion of Theorem
 \ref{th:onepert}.\ \ \ []

\section{Diffusion of power
 in discrete frequency (nonradiative) systems}\label{se:ex}

In the previous section we calculated the effect of a single defect on
the the mode-power distribution.
In this section we show how to apply this result to prove diffusion
of power for the perturbed Hamiltonian
system, (\ref{eq:perturbed}), where $g(t)$ is a random function of the form
 (\ref{eq:gdef}), defined in terms of a random sequence $\{d_j\}_{j\ge0}$.
In particular, the sequence $\{d_j\}_{j\ge0}$ will be taken to be generated
 by independent, identically distributed random variables.
This will be result in a {\it mixing the phases} of the complex mode amplitudes,
 after each defect.

We assume that {\bf (H1-H3)} are satisfied. The following
hypothesis ensures that (\ref{eq:IC}) holds before each
defect, thus enabling repeated application of Theorem \ref{th:onepert}.

{\bf (H4)} $d_0,d_1,\ldots$ are independent identically distributed random
variables taking only nonnegative values and such that for any
$l\in\{0,1,\ldots \}$ and $j\neq k\in\{1,2\ldots \}$ we have
$$\E\left(e^{i(\la_j-\la_k)d_l}\right)=0$$
where $\E ( \cdot )$ denotes the expected value.

 Clearly  (H4) requires the eigenvalues to be distinct but aside from these we
claim that for any {\em finitely many, distinct} eigenvalues
$\la_1,\la_2,\ldots ,\la_m$ there exist a random variable
satisfying (H4).

\nit {\bf Example 1 (finitely many bound states)} Given
$\la_1,\la_2,\ldots,\la_m$ distinct choose the random variables
$d_l,\ l=0,1,\ldots $ to be identically distributed with distribution $d:$
$$d=\sum_{1\le j<k\le m}d_{jk}$$
where $d_{jk}$ are independent random variables such that the distribution
of $d_{jk}$ is uniform on the interval $[0,2\pi/|\la_j-\la_k|]$. In this case,
 for any $j'\neq k'\in\{1,2,\ldots \}$
\ba
\E\left(e^{i(\la_{j'}-\la_{k'})d}\right)&=&
\E\left(\prod_{1\le j<k}e^{i(\la_{j'}-\la_{k'})d_{jk}}\right)\no\\
&=&\prod_{1\le j<k}\E\left( e^{i(\la_{j'}-\la_{k'})d_{jk}}\right)\no\\
&=& 0\no
\ea
since $\E\left(e^{i(\la_{j'}-\la_{k'})d_{j'k'}}\right)=0$.

Another choice is to consider discrete $d_{jk}$'s. Namely, take
$d_{jk}$ to be the discrete random variable taking each of the
values $0$ and $\pi/|\la_j-\la_k|$ with probability $1/2$. A
concrete example is, in the case we have three eigenvalues
$\la_1<\la_2<\la_3$, to choose $d$ to be the random variable
taking each of the eight values:
$$0,\frac{\pi}{\la_2-\la_1},\frac{\pi}{\la_3-\la_1},\frac{\pi}{\la_3-\la_2},$$
$$\pi\left(\frac{1}{\la_2-\la_1}+\frac{1}{\la_3-\la_1}\right),
\pi\left(\frac{1}{\la_2-\la_1}+\frac{1}{\la_3-\la_2}\right),
\pi\left(\frac{1}{\la_3-\la_1}+\frac{1}{\la_3-\la_2}\right),$$
$$\pi\left(\frac{1}{\la_2-\la_1}+\frac{1}{\la_3-\la_1}+\frac{1}{\la_3-\la_2}\right)$$
with probability $1/8$.

(H4) does not restrict us to system with finitely many bound states:

\nit {\bf Example 2 (infinitely many bound states)} Let the quantum harmonic
oscillator in one dimension:
$$H_0=-{\hbar^2\over 2}\partial_x^2+\omega^2x^2,\qquad x\in\R,$$
be the unperturbed Hamiltonian. Then
$$ \la_n=\hbar\omega (n+1/2),\qquad n=0,1,2,\ldots ,$$
see for example \cite{kn:L-L}. Note that (H4) holds provided that we choose
$d_l,\ l=0,1,\ldots $ to be identically and uniformly distributed on the
interval $[0,2\pi/(\hbar\omega)].$

\un{Note on degenerate eigenvalues:} As discussed above (H4)
cannot be satisfied in the case $H_0$ admits degenerate
eigenvalues. However, at least in some cases, our theory can be applied. In general the
degeneracy is a consequence of the symmetries of $H_0$, i.e. the
existence of a self-adjoint operator, say $L$, commuting with
$H_0$, $[L,H_0]=0$. To recover our results it is sufficient to
assume that $\beta $, the ``space-like" part of the perturbation,
respects the symmetry, i.e. commutes with $L$. One
can now factor out $L ,$ i.e. work on the invariant subspaces of $L$
where $H_0$ is nondegenerate. Along the lines of Example 2 one can
consider the quantum harmonic oscillator in three
dimensions which has a spherically symmetric Hamiltonian and degenerate
eigenvalues, see for example \cite{kn:L-L}. If $\beta$ is
spherically symmetric then it only couples bound states with
the same angular momentum. Hence the problem reduces to subsystems consisting
of bound states with the same angular momentum but different energy, therefore
nondegenerate. The choice we made in Example 2 will satisfy (H4) in each
of the subsystems.

\subsection{Power diffusion after a fixed (large) number of defects}

\begin{theo}\label{th:train} Consider equation (\ref{eq:schrodinger}) with $g$
of the form (\ref{eq:gdef}). Assume {\bf (H1-H4)} hold. Then the
expected value of the power vector after passing a fixed number of
perturbations $``n"$ satisfies
\be\label{eq:train}P^{(n)}=T_\ve^nP(0)+\cO (n\ve^3),\ee
where $T_\ve$ is given in (\ref{eq:nTdef})
\ba
&& P_k^{(n)}=\E(|a_k(t)|^2),\ k=1,2,\ldots\label{eq:Pkn-def}\\
&& t_{n-1}+T\le t\le t_n,\ (t\ {\rm ranging\ between\ the}\
n^{th}\
 {\rm and}\ (n+1)^{st}\
 {\rm defects}
\nn
\ea
\end{theo}

\nit\un{Proof.} We will prove the theorem by induction
on $n\ge 0$, the number of defects traversed.
 For $n=0$ the assertion is obvious.
Suppose now that for  $n\ge 0$  we have
\be\label{eq:indhyp}P^{(n)}=T_\ve^nP(0)+\cO (n\ve^3) .\ee
 We will show
\be
\label{eq:conchyp}
 P^{(n+1)}=T_\ve^{n+1}P(0)+\cO\left((n+1)\ve^3\right)
\ee
by applying Theorem \ref{th:onepert} to (\ref{eq:indhyp}).
 In order to apply Theorem \ref{th:onepert} we need to verify
that (\ref{eq:IC}) is satisfied before the $n+1^{{\rm st}}$
defect. Specifically, we must verify that
 for any pair $k\neq j$
\be
\label{eq:ICnp1}
\E\left(a_k(t_{n+1})\bar a_j(t_{n+1}) \right)\ =\
 \E\left(a_k \left(nT+{\textstyle\sum_{k=0}^{n+1}}d_k\right)
\overline a_j\left(nT+\textstyle{\sum_{k=0}^{n+1}}
d_k\right)\right)=0. \ee
 Using
the fact that $d_{n+1}$ is independent of $d_0+d_1+\ldots +d_n,$
and {\bf (H4)} we have: \ba \E\left(a_k\overline
a_j\left(nT+{\textstyle\sum_{k=0}^{n+1}}d_k\right)\right)&=&
\E\left(a_k\overline
a_j\left(nT+{\textstyle\sum_{k=0}^n}d_k\right)e^{i(\la_j-\la_k)d_{n+1}}\right)\no\\
&=&\E\left(a_k\overline
a_j\left(nT+{\textstyle\sum_{k=0}^n}d_k\right)\right)
\E\left(e^{i(\la_j-\la_k)d_{n+1}}\right)=0.\no \ea Thus
(\ref{eq:ICnp1}) holds and all the hypothesis of Theorem
\ref{th:onepert} are now satisfied. By applying it and using
(\ref{eq:indhyp}) we have \ba
P^{(n+1)}&=&T_\ve P^{(n)}+\cO (\ve^3)\no\\
&=&T_\ve\left(T_\ve^nP(0)+\cO( n\ve^3)\right)+\cO (\ve^3)\no\\
&=&T_\ve^{n+1}P(0)+\cO ((n+1)\ve^3).\no \ea
Hence (\ref{eq:indhyp}) implies (\ref{eq:conchyp}). This concludes the
induction step and the proof of Theorem \ref{th:train} is now complete.\ \ \ []

In the next two Corollaries we describe the asymptotic behavior of
the vector of expected powers when the number of defects $n$ tends to
infinity. Note that after a possible reordering of the
eigenvectors $\psi_1,\psi_2,\ldots ,$ of $H_0,$ the operator $B$
given by (\ref{eq:rBdef}) might look like\footnote{For such a
decomposition to occur it is sufficient that $H_0$ and $\beta$
have common invariant subspaces ${\cal H}_1\subset {\cal H}, {\cal
H}_2\subset {\cal H},\ldots ,{\cal H}_q\subset {\cal H},\ldots$}:
\be\label{eq:reduce} B=\diag\left[B_1,B_2,\ldots
,B_q,\ldots\right],\ee where $B_1, B_2,\ldots ,B_q ,\ldots $ are
square matrices (linear operators) of dimensions $m_1, m_2,\ldots
,m_q ,\ldots ,$ $1\le m_q \le\infty,\ q=1,2,\ldots .$ In linear
algebra terms this means that $B$ is reducible. In terms of the
dynamical system (\ref{eq:train}) generated by
$T_\ve=\mathbb{I}-\ve^2B$ it means that, after a possible
reordering, the first $m_1$ bound states of $H_0$ are isolated
from the rest. The same is valid for the next $m_2$ bound states, etc.
To understand the evolution of the full system it is sufficient to
analyze each of the isolated subsystems separately. They all
evolve according to (\ref{eq:train}) with
$T_\ve=\mathbb{I}-\ve^2B_q$ and $B_q$ given by (\ref{eq:rBdef})
but the indices span only a subset of the eigenvectors
$\psi_1,\psi_2,\ldots $ of $H_0.$ The main difference is that now
$B_q$ is irreducible. In what follows we are focusing on one such
subsystem and drop the index $q.$

\begin{cor}\label{rmk:resultinterpretfinite}
 If the subsystem has a finite number of
bound states, say $m,$ then
\be\label{eq:Blimitfinite}
\lim_{n\rightarrow \infty}P^{(n)}=\left\{\begin{array}{lr}
P(0),&{\rm if}\ n\ll\ve^{-2}\\
e^{-B\tau}P(0) & {\rm if}\ n=\tau\ve^{-2}\\
{E\over m}(1,1,\ldots ,1)',&{\rm if}\ \ve^{-2}\ll n\ll |\ve |^{-3}
 \end{array}\right. ,
\ee where $E=P_1(0)+P_2(0)+\ldots +P_m(0)$ is the expected total
power in the subsystem and it is conserved.
\end{cor}

\nit\un{Proof.} We use the following properties of the irreducible
matrix $B$:
\begin{enumerate}
\item[(B1)] $B$ is self adjoint and $B\ge 0;$
\item[(B2)] $0$ is a simple eigenvalue for $B$ with corresponding
normalized eigenvector
\be
r_0=\frac{1}{\sqrt{m}}\left(1,1,\ldots ,1\right)'.
\label{eq:r0def}
\ee
\end{enumerate}
These properties are proved in the Appendix.

Let $\beta_0=0,\beta_1,\beta_2,\ldots ,\beta_{m-1}$ be the
eigenvalues of $B$ counting multiplicity, and let $r_0,r_1,\ldots
,r_{m-1}$ be the corresponding orthonormalized eigenvectors. By (B1)
and (B2) $\beta_1,\beta_2,\ldots ,\beta_{m-1}$ are strictly
positive. Let
$$ R=\left[r_0,r_1,\ldots ,r_{m-1}\right]$$
be the matrix whose columns are orthonormalized eigenvectors of $B$
 and let $R'$ be its transpose. Then
 \ba
R'BR&=&\diag\left[\beta_0,\beta_1,\beta_2,\ldots ,\beta_{m-1}\right]\no\\
R'R&=&\mathbb{I}\ =\ RR'.\no \ea
It follows that
 \ba T_\ve^n &=&
\left(\mathbb{I}-\ve^2B\right)^n=R\left[R'\left(\mathbb{I}-\ve^2B\right)R\right]^nR'\no\\
&=&R\diag\left[(1-\ve^2\beta_0)^n,(1-\ve^2\beta_1)^n,\ldots
,(1-\ve^2\beta_{m-1})^n\right]R'.\no \ea
We now study $\lim_{n\to\infty} T_\ve^n$ for the three asymptotic regimes of
  (\ref{eq:Blimitfinite}).
Note that for $0\le k\le m-1$ we have:
\ba
\lim_{n\to\infty, \ve^2n\to0} (1-\ve^2\beta_k)^n\ &=&\ 1\nn\\
\lim_{n\to\infty, \ve^2n=\tau}(1-\ve^2\beta_k)^n \ &=&\ e^{-\beta_k\tau}\nn\\
\lim_{n\to\infty, \ve^2n\to\infty} (1-\ve^2\beta_k)^n\
  &=&\ 0,\ \beta_k>0\nn\\
\lim_{n\to\infty, \ve^2n\to\infty} (1-\ve^2\beta_k)^n\
  &=&\ 1,\ \beta_k=0
\label{eq:limits}
\ea
Consequently,
\be\label{eq:Tepsnlim}
\lim_{n\rightarrow\infty}T_\ve^n=
\left\{
\begin{array}{lr} R\diag[1,1,\ldots ,1]R'=\mathbb{I}& {\rm if}\ n\ll\ve^2 \\
R\diag\left[e^{-\beta_0\tau},e^{-\beta_1\tau},\ldots ,e^{-\beta_{m-1}\tau}\right]=e^{-B\tau}& {\rm if}\ n=\tau\ve^{-2} \\
R\diag[1,0,0,\ldots ,0]R'={\rm projection\ onto\ }r_0& {\rm if}\
\ve^{-2}\ll n\ll |\ve |^{-3}
\end{array}\right. ,
\ee
where $r_0$ is defined in (\ref{eq:r0def}).

Substitution of  (\ref{eq:Tepsnlim}) into (\ref{eq:train}) completes the
proof of Corollary \ref{rmk:resultinterpretfinite}.\ \ \ []

\begin{cor}\label{rmk:resultinterpretinfinite}
 If the subsystem has an infinite number
of bound states, then
\be\label{eq:Blimitinfinite}
\lim_{n\rightarrow 0}P^{(n)}=\left\{\begin{array}{lr}
P(0),&{\rm if}\ n\ll\ve^{-2}\\
e^{-B\tau}P(0) & {\rm if}\ n=\tau\ve^{-2}
 \end{array}\right. ,
\ee For $n\gg\ve^{-2}$ the limit in $\ell^2$ is 0, while the limit
in $\ell^1$ does not exist. More precisely, although the total
power in the subsystem is conserved, \be \sum_{k=1}^\infty
P_k^{(n)}\ =\ E,\qquad \forall n\ge 0, \nn\ee $\{P^{(n)}\}$ does
not converge in $\ell^1$ due to an energy transfer to the high
modes. In particular, for any fixed $N\ge1$: \ba \lim_{n\to\infty}
 \sum_{k=N}^\infty P_k^{(n)}\ &=&\ E,\nn\\
\lim_{n\to\infty}\sum_{k=1}^N P_k^{(n)}\ &=&\ 0\label{eq:flux}.
\ea
 \end{cor}
We note that similar results have been obtained in \cite{leb:diff}
but for different types of random perturbation.

Corollaries \ref{rmk:resultinterpretfinite} and \ref{rmk:resultinterpretinfinite}
 show that, on time scales of order
$1/\ve^2$, the dynamical system is equivalent with
\be\label{eq:heat}\partial_\tau P(\tau)=-BP(\tau).\ee
 Moreover the
definition of $-B$ in (\ref{eq:rBdef}) together with $-B\le 0$ and
$\ e^{-B}$ unitary on $\ell^1$ implies that the flow (\ref{eq:heat})
is very much like that of a discrete heat or diffusion equation.

In conclusion the number of defects encountered should be comparable with
  $1/\ve^2$ to have
a significant effect. Once they are numerous enough, the defects diffuse the
power in the system.
If the number of defects is much larger than $1/\ve^2$  the power becomes
uniformly distributed among the bound states.

\begin{rmk} Hyptothesis {\bf (H4)} is important.
  If we do not assume {\bf (H4)} then
 the correction term for each defect is of size
$\ve,$ since the boundary terms (\ref{eq:bt}) no longer vanish.
Consequently the correction term in the main result
(\ref{eq:train}) is $\cO (n\ve)$ which on the ``diffusion time
scale" $n\sim\ve^{-2}$ is very large.
\end{rmk}

\nit\un{Proof of Corollary \ref{rmk:resultinterpretinfinite}} In
the case of an infinite number of bound states $B$
 has the following properties, see the Appendix:
\begin{enumerate}
\item[(B1$_\infty$)] $B$ is a nonnegative, bounded self adjoint operator
on $\ell^2$ with spectral radius less or equal to $2;$
\item[(B2$_\infty$)] $0$ is not an eigenvalue for $B;$
\item[(B3$_\infty$)] $B$ is a bounded operator on $\ell^1$ with norm $\|B\|_1\le 2;$
\item[(B4$_\infty$)] For $|\ve|\le 1$ the operator
$T_\ve=\left(\mathbb{I}-\ve^2B\right)$ transforms positive vectors
(i.e. all components positive) into positive vectors and conserves
their $\ell^1$ norm.
\end{enumerate}

We are going to focus first on $\ell^2$ results. Based on the
spectral representation theorem, see \cite{kn:RS1}, we have for
any Borel measurable real function $f:$ \be\label{eq:specrep}
f(B)=\int_0^2f(s)d\mu(s).\ee Here $d\mu(s)$ is the spectral
measure induced by $B$. Note that B2$_\infty$ implies the continuity of
$\mu(s) $ at zero.

Now
$$T_\ve^n=\left(\mathbb{I}-\ve^2B\right)^n=\int_0^2\left(1-\ve^2s\right)^nd\mu(s)$$
and \be\label{eq:specTeps}
\lim_{n\rightarrow\infty}T_\ve^n=\lim_{\ve\rightarrow
0}\int_0^2\left(1-\ve^2s\right)^nd\mu(s)=\int_0^2\lim_{\ve\rightarrow
0} \left(1-\ve^2s\right)^nd\mu(s).\ee For the last equality we
used the dominant convergence theorem with $|1-\ve^2s|^n\le 1$ for
$0\le s\le 2,\ |\ve |\le 1$ and $\int_0^21d\mu(s)=\mathbb{I}.$
Using (\ref{eq:limits}), with $s$ replacing $\beta_k$, we have
that (\ref{eq:specTeps}) becomes
\be\label{eq:iTepsnlim}
\lim_{n\rightarrow\infty}T_\ve^n=\left\{\begin{array}{lr} \int_0^2 1d\mu(s)=\mathbb{I}& {\rm if}\ n\ll\ve^2 \\
\int_0^2 e^{-s\tau}d\mu(s)=e^{-B\tau}& {\rm if}\ n=\tau\ve^{-2} \\
\mu(0+)-\mu(0)=0& {\rm if}\ \ve^{-2}\ll n\ll |\ve |^{-3}
\end{array}\right.,\ee
where we used (\ref{eq:specrep}) and the continuity of $\mu(s)$ at
zero .

Plugging (\ref{eq:iTepsnlim}) in (\ref{eq:train}) gives the
required results in $\ell^2.$

For the results in $\ell^1$ we use series expansions:
\be\label{eq:seriesexp}
 \left(\mathbb{I}-\ve^2B\right)^n=\mathbb{I}+\left(\begin{array}{c}
n\\ 1\end{array}\right)\ve^2(-B)+\left(\begin{array}{c} n\\
2\end{array}\right)\ve^4(-B)^2+\ldots
+\left(\begin{array}{c} n\\
n\end{array}\right)\ve^{2n}(-B)^n\ee
Since $\|B\|_1\le 2,$ (see
property B3$_\infty$), the finite series above is dominated in $\ell^1$
operator norm by:
\be\label{eq:seriesest} 1+2\ve^2\left(\begin{array}{c} n\\
1\end{array}\right)+(2\ve^2)^2\left(\begin{array}{c} n\\
2\end{array}\right)+\ldots +(2\ve^2)^n\left(\begin{array}{c} n\\
n\end{array}\right)=(1+2\ve^2)^n\le e^{2n\ve^2}.\ee

As $n\rightarrow\infty$ the series in
(\ref{eq:seriesest}) becomes infinite. However, as long as
$n\le\tau/\ve^2,\ \tau>0$ fixed, the sum in (\ref{eq:seriesest})
 is finite and  hence
that in (\ref{eq:seriesexp}) is convergent. Now for each $k=1,2,\ldots $
the $(k+1)^{{\rm st}}$ term in the series (\ref{eq:seriesexp}) has
the property:
$$\lim_{n\rightarrow\infty}\left(\begin{array}{c}
n\\ k\end{array}\right)\ve^{2k}(-B)^k=\left\{\begin{array}{lr}
0&{\rm if}\ n\ll\ve^{-1}\\ \frac{\tau^k}{k!}(-B)^k&{\rm if}\
n=\tau\ve^{-2}\end{array}\right. $$ Hence by the Weierstrass
criterion for absolutely convergent series we have:
\be\label{eq:l1Tepslim}
\lim_{n\rightarrow\infty}T_\ve^n=\lim_{n\rightarrow\infty}\left(\mathbb{I}-\ve^2B\right)^n
=\left\{\begin{array}{lr} \mathbb{I}-0+0-\ldots =\mathbb{I}&{\rm if}\ n\ll\ve^{-1}\\
\mathbb{I}-\tau B+\frac{(\tau B) ^2}{2!}-\frac{(\tau B)
^3}{3!}+\ldots =e^{-\tau B} &{\rm if}\
n=\tau\ve^{-2}\end{array}\right.\ee

It remains to prove that as $n\to\infty, \ve^2n\to\infty$, $\{P^{(n)}\}$
 does not converge in  $\ell^1$.
 Let $P^{(0)}\in \ell^1\cap \ell^2$  denote a vector with positive components,
 and consider the sequence:
\be P^{(n)}=T_\ve^nP^{(0)}\in \ell^1\cap \ell^2.
\label{eq:Pnseq}
\ee
By the third part of (\ref{eq:iTepsnlim}),
$\|P^{(n)}\|_2\to0$.  Assume now that there
exists $P\in \ell^1$ such that
$\|P^{(n)}-P\|_1=0$.  Since both $\ell^1$ and
$\ell^2$ convergence imply convergence of each component,
we deduce that $P=0$.
On the other hand,
by $P^{(n)}=T_\ve P^{(n-1)},\qquad n=1,2,\ldots$
 and property B4$_\infty$, we deduce that $P^{(n)}$ is a positive
vector for which  $\|P^{(n)}\|_1=\|P^{(0)}\|_1\stackrel{{\scriptstyle
def}}{=} E>0$ for all $n\ge 0.$ Consequently $P$ is a nonnegative
vector with $\|P\|_1=E>0$, a contradiction.
The proof of the Corollary is now complete. \ \ \ [].

\subsection{Power diffusion after a fixed (large) time interval and a random
number of defects}

As pointed out in its statement, Theorem \ref{th:train} is valid
when one measures the power vector after a fixed number of defects
$``n"$ regardless of the realizations of the random variables.
That is after each realization of $d_0,d_1,\ldots $ the power
vector is measured in between the $n^{{\rm th}}$ and the
$(n+1)^{{\rm st}}$ defect. Averaging the measurements over all the
realizations of $d_0,d_1,d_2,\ldots $ gives the result of Theorem
\ref{th:train}. What happens if one chooses to measure the power
vector at a fixed time $``t"$ (i.e. a fixed distance along the
fiber)? The answer is given by the next theorem:

\begin{theo}\label{th:fixedtime} Consider equation (\ref{eq:schrodinger}) with
$g$ of the form (\ref{eq:gdef}). Assume that {\bf (H1-H4)} are
satisfied and that all random variables $d_0,d_1,\ldots ,$ have
finite mean, variance and third momentum. Fix a time $t,\ 0\le
t\ll 1/|\ve |^3$. Then the expected value of the power vector at a
fixed time $P(t)$ satisfies \be P(t)=T_\ve^nP(0)+\cO
(\max\{t\ve^3, \ve^{4/5}\}), \label{eq:Pt} \ee where $n=\lfloor
t/(T+M)\rfloor$ denotes the integer part of $t/(T+M),\ T$ is the
common time span of the defects and $M$ is the mean of the
identically distributed random variables $d_0,d_1,\ldots .$
\end{theo}

\begin{cor} In this setting, the conclusions of Corollaries
\ref{rmk:resultinterpretfinite}, \ref{rmk:resultinterpretinfinite}
hold with $n$ replaced by $t.$
\end{cor}

\nit\un{Proof of Theorem \ref{th:fixedtime}.} As before, let
$P^{(k)}$ be the expected power vector after exactly $``k"$
defects. Denote by $N$ the random variable counting the number of
``defects" up until time $t$, i.e.
\be\label{eq:Ndef}(N-1)T+d_0+\ldots +d_{N-1}<t\le NT+d_0+\ldots
+d_N.\ee and let $\delta(\ve)$ denote the integer, which grows as $\ve$
decreases:
 \ba
\tilde\delta
&=&\max\left\{1.39\left(\frac{\rho}{\sigma^2(T+M)}\right)^{2/5}\ve^{-6/5},
\frac{\sigma}{T+M}\sqrt{n\log\left(\ve^{-2}\right)}+
\left(\frac{\sigma}{T+M}\right)^2\log\left(\ve^{-2}\right)\right\}\no\\
\delta &=&\lfloor\tilde\delta\rfloor +1,\label{eq:variance}\ea
where $M ,\sigma^2 ,$ respectively $\rho$ are the mean, variance
and the centered third momentum, of the identically distributed
variables $d_0,d_1,d_2,\ldots ,$ and $n$ is the integer part of
$t/(T+M).$ Note that for $t\sim\ve^{-3}$ or smaller
$\delta\ll\ve^{-2} .$ The choice of $\delta(\ve)$ is explained
below.

The proof consists of three stages:
\begin{enumerate}
\item $P(t)=P^{(n+\delta)}+\cO (\ve)+\cO (\delta\ve^2)$
\item $P^{(n+\delta)}=P^{(n)}+\cO (\delta\ve^2)$
\item $P^{(n)}=T_\ve^nP(0)+\cO (n\ve^3)$
\end{enumerate}
where $n=\lfloor t/(T+M)\rfloor .$ The last stage is simply
Theorem \ref{th:train}.

For the second stage one applies again the previous theorem to get:
$$P^{(n+\delta)}=T_\ve^\delta P^{(n)}+\cO (\delta\ve^3).$$
Now $T_\ve=I-\cO (\ve^2)$ and since $\delta\ll \ve^{-2}$ stage two follows.

The first stage is the trickiest. Without loss of generality we
can assume that $t/(T+M)$ is an integer. Indeed, for $n=\lfloor
t/(T+M)\rfloor $ we have
$$P(t)-P(n(T+d))=\cO \left(\ve (T+M)\right)=\cO(\ve),$$
an error which is already accounted for in this stage.

Suppose first $n-\delta\le N\le n+\delta ,$ i.e. we condition the expected values
to the realization of $|N-\delta|\le 0.$ Then the difference
between the conditional expected values of the power vector at
time $t$ and after $n+\delta$ defects is of order $\cO (\ve)+\cO
(\delta\ve^2)$. This follows from the fact that the condition
$n-\delta\le N\le n+\delta$ restricts only the realizations of
$d_0,d_1,\ldots d_N$ leaving the realizations of $d_{N+1},\ldots
d_{n+\delta}$ arbitrary; see (\ref{eq:Ndef}). Hence, as in stage
two, the conditional expected values satisfy:
$$P^{(n+\delta)}=P^{(N+1)}+\cO (\delta\ve^2).$$
In addition
$$P^{(N+1)}=P(t)+\cO (\ve),$$
since there are at most 2 defects of size $\ve$ from $``t"$ up until after the
$(N+1)^{th}$ defect.

Let $p(t)$ denote the power vector
$$p(t)=\left(|a_1(t)|^2,|a_2(t)|^2,\ldots\right).$$ Recall that
by definition $P(t)=\E (p(t))$ and the total power in the system
(\ref{eq:schrodinger}) is conserved, {\em i.e.}
$$\|p(t)\|_1\stackrel{\scriptstyle{{\rm
def}}}{=}\sum_k|a_k(t)|^2\equiv \|p(0)\|_1,\quad t\in\R .$$
Moreover, \ba P(t)&=&\E\left(p(t)\ :\ |N-n|\le\delta\right)+
\E\left(p(t)\ :\ |N-n|>\delta\right)\no\\
&=&P^{(n+\delta)}+\cO(\delta\ve^2)+\cO(\ve)+\cO\left(\|p(0)\|_1{\rm
Prob}(|N-n|>\delta)\right)\label{eq:stage1}\ea We claim that for
$\delta$ given by (\ref{eq:variance}) \be\label{eq:probvar} {\rm
Prob}(|N-n|>\delta)=\cO(\ve)+\cO(\delta\ve^2).\ee Indeed, since
$t=n(T+M)$ \ba {\rm Prob}(|N-n|>\delta)&=&{\rm
Prob}\left(\sum_{k=0}^{n+\delta}(T+d_k)\le
t\right)+{\rm Prob}\left(\sum_{k=0}^{n-\delta}(T+d_k)>t\right)\no\\
&=&{\rm Prob}\left(\frac{\sum_{k=0}^{n+\delta }(T+d_k)-(n+\delta
)(T+M)}{\sigma\sqrt{n+\delta
}}\le\ -\ \frac{\delta (T+M)}{\sigma\sqrt{n+\delta }}\right)\no\\
&+&{\rm Prob}\left(\frac{\sum_{k=0}^{n-\delta}(T+d_k)-(n-\delta
)(T+M)}{\sigma\sqrt{n-\delta }}
>\frac{\delta (T+M)}{\sigma\sqrt{n-\delta }}\right).
\label{eq:cltid}\ea

 We are going to show how the choice
(\ref{eq:variance}) implies \be\label{eq:cltid1} {\rm
Prob}\left(\frac{\sum_{k=0}^{n-\delta}(T+d_k)-(n-\delta
)(T+M)}{\sigma\sqrt{n-\delta }}
>\frac{\delta (T+M)}{\sigma\sqrt{n-\delta }}\right)\le {\ve\over 2}+{\delta\ve^2\over
2}.\ee The other half of (\ref{eq:cltid}): \be\label{eq:cltid2}
{\rm Prob}\left(\frac{\sum_{k=0}^{n+\delta}(T+d_k)-(n+\delta
)(T+M)}{\sigma\sqrt{n+\delta }} <-\frac{\delta
(T+M)}{\sigma\sqrt{n+\delta }}\right)\le {\ve\over
2}+{\delta\ve^2\over 2}.\ee
 is analogous.

Depending on the size of $n$ one has either: \be\label{eq:rest1}
\frac{0.8\rho}{\sigma^3\sqrt{n-\delta}}\le {\delta\ve^2\over 2}\ee
or: \be\label{eq:rest2}\frac{0.8\rho}{\sigma^3\sqrt{n-\delta}}>
{\delta\ve^2\over 2}.\ee If (\ref{eq:rest1}) holds, which
corresponds to large $n,$ we use the central limit theorem with
Van Beek rate of convergence, see \cite{dur:pro}:
$${\rm Prob}\left(\frac{\sum_{k=0}^{n-\delta}(T+d_k)-(n-\delta
)(T+M)}{\sigma\sqrt{n-\delta }}
>\frac{\delta (T+M)}{\sigma\sqrt{n-\delta }}\right)\le {1\over\sqrt{2\pi}}\int_{\frac{\delta (T+M)}{\sigma\sqrt{n-\delta
}}}^\infty e^{-x^2/2} dx + \frac{0.8\rho}{\sigma^3\sqrt{n-\delta}}
.$$ This together with (\ref{eq:rest1}), the inequality
$${1\over\sqrt{2\pi}}\int_a^\infty e^{-x^2/2} dx\le
\frac{e^{-a^2/2}}{2}$$ and the fact that $\delta\ge {\sigma\over
(T+M)}\sqrt{n\log\ve^{-2}}$ implies $\frac{\delta
(T+M)}{\sigma\sqrt{n-\delta }}\ge 2\ln\ve^{-1},$ proves
(\ref{eq:cltid1}) for the case (\ref{eq:rest1}). If
(\ref{eq:rest2}) holds then we apply Chebyshev inequality:
$${\rm Prob}\left(\frac{\sum_{k=0}^{n-\delta}(T+d_k)-(n-\delta
)(T+M)}{\sigma\sqrt{n-\delta }}
>\frac{\delta (T+M)}{\sigma\sqrt{n-\delta }}\right)\le
\frac{\sigma^2(n-\delta )}{\delta^2 (T+M)^2}\le {\delta\ve^2\over
2},$$ where the latter inequality follows from (\ref{eq:rest2})
and $$\delta\le
1.39\left(\frac{\rho}{\sigma^2(T+M)}\right)^{2/5}\ve^{-6/5}.$$

 From (\ref{eq:cltid}), (\ref{eq:cltid1}) and (\ref{eq:cltid2}) we
get relation (\ref{eq:probvar}). The latter plugged into
(\ref{eq:stage1}) proves the first stage.

Finally, the three stages imply Theorem \ref{th:fixedtime}
provided that both $\ve$ and $\delta\ve^2$ are dominated by
$C\max\{n\ve^3, \ve^{3/4}\},$ for an appropriate constant $C>0.$
This follows directly from $\ve\le 1$ and (\ref{eq:variance}). The
proof is now complete.\ \ \ []

\section{Diffusion of power in systems with
discrete and continuous spectrum
 }\label{se:yesrad}

Thusfar we have considered with systems with Hamiltonian, $H_0$, having
only discrete spectrum. We now extend our analysis to the case where $H_0$
has both discrete and continuous spectrum. Continuous spectrum is associated
with radiative behavior and this is manifested in a {\it dissipative}
 correction to the operator (\ref{eq:nTdef}), entering at ${\cal O}(\ve^2)$.
Therefore, the dynamics on time scales $n\sim\ve^{-2}$ is characterized
by diffusion of energy among the discrete modes {\bf and} radiative damping
due to coupling of bound modes to the ``heat bath'' of radiation modes.

The hypotheses on the unperturbed Hamiltonian $H_0$ are similar to
those in \cite{kw:mbs}. There is one exception though, the
singular local decay estimates are replaced by a condition
appropriate for perturbations with continuous spectral components,
see Hypothesis {\bf (H7')} below. For convenience we list here and
label all the hypotheses we use:

{\bf (H1')} $H_0$ is self-adjoint on the Hilbert space ${\cal
H}$. The norm, respectively scalar product, on ${\cal H}$ are
denoted by $\|\cdot\|$, respectively $\lan \cdot ,\cdot\ran .$

{\bf (H2')} The spectrum of  $H_0$  is assumed to consist of
an absolutely continuous part,
 $\sigma_{\rm cont}(H_0)$, with associated spectral projection,
 $\Pc$, spectral measure $\dm(\xi)$ and a discrete part formed by isolated eigenvalues
 $\lambda_1,\lambda_2,\dots ,\lambda_m$ (counting
multiplicity)
 with an orthonormalized set of
 eigenvectors $\psi_1,\psi_2,\dots ,\psi_m$,
 i.e.
for $k,\ j=1,\dots,m$
$$
 H_0\psi_k=\lambda_k \psi_k,\
 \lan\psi_k,\psi_j\ran=\delta_{kj},
$$ where $\delta_{kj}$ is the Kronecker-delta symbol.

 {\bf (H3')} \underline{Local decay estimates on
$e^{-iH_0t}$}: There exist self-adjoint "weights", $\wmi$, $\wpl$,
number $r_1>1$
 and a constant $\cal C$ such that

\ (i)\ $w_+$ is defined on a dense subspace of ${\cal H}$ and on
 which $w_+\ge cI$, $c > 0$

\ (ii)\ $w_-$ is bounded, i.e. $w_- \in {\cal L}({\cal H}),$ such that
$Range(\wmi)\subseteq Domain(\wpl)$

\ (iii)\ $\wpl\ \wmi\ \Pc\ =\ \Pc $ and $\Pc\ =\ \Pc\ \wmi\ \wpl$ on the domain
of $\wpl$

\nit and for all $f\in{\cal H}$ satisfying $\wpl f\in{\cal H}$ we have
$$\|  \wmi e^{-iH_0t} {\bf P_c} f \|\le {\cal C}\ \lan t\ran^{-r_1}
\|\wpl f\|,\ \ t\in\R. $$

The hypotheses on the perturbation are similar to the ones used in
the previous sections for discrete systems, namely:

 {\bf (H4')} $\beta$ is a bounded self adjoint operator on
${\cal H}$ and satisfies $\|\beta\|=1$. In addition we suppose
that $\beta$ is ``localized", i.e. $\wpl\beta $ and
$\wpl\beta\wpl$ are bounded on ${\cal H},$ respectively on
$Domain(\wpl).$

 {\bf (H5')} $g_0(t)\in L^2(\R)$ is real valued, has compact
support contained in $[0,T]$ on the positive real line and its
$L^1$-norm, denoted by $\|g_0\|_1$ is $1$. Therefore its Fourier
transform, $\hat g_0$ is smooth and $\|\hat g_0\|_\infty\le1$.

 {\bf (H6')} $d_0,d_1,\ldots$ are independent identically
distributed random variables taking only nonnegative values, with
finite mean, $M,$ and such that for any $l\in\{0,1,\ldots \}$ and
$j\neq k\in\{1,2\ldots ,m \}$ we have
$$\E\left( e^{i(\la_j-\la_k)d_l}\right)=0$$
where $\E\left( \cdot\right)$ denotes the expected value.

Define the common characteristic (moment generating) function for
the random variables $d_0+T,d_1+T,\ldots $
\be\label{eq:generating} \rho(\xi)\ \equiv\ \E\left(
e^{-i\xi(d_0+T)}\right)=\E\left( e^{-i\xi(d_1+T)}\right)=\cdots
.\ee Note that $\rho$ is a continuous function on $\R$ bounded by
$1$. Then (H6') is equivalent to
$$\rho(\la_k-\la_j)=0$$
for all $j\neq k\in\{1,2,\ldots, m\}.$

We require an additional local decay estimate:

 {\bf (H7')} There exists the number $r_2>2$ such that for all
$f\in{\cal H}$ satisfying $\wpl f\in{\cal H}$ and all $\la_k,\ \la_j,\
k,j=1,\dots,m $ we have:
$$\|  \wmi e^{-iH_0t}\rho(H_0-\la_k)\hat g_0(H_0-\la_k)\hat
 g_0(\la_j-H_0) {\bf P_c} f \|
\le\frac{{\cal C}\|g_0\|_1^2}{\lan t\ran^{r_2}} \|\wpl f\|,\ \
t\in\R.
$$
Here $\hat g_0$ denotes the Fourier Transform, see
(\ref{eq:FT}), and the operators $\rho(H_0-\la)\Pc,\ \hat
g_0(\la-H_0)\Pc$ are defined via the spectral theorem:
 \ba
\rho(H_0-\la)\Pc&=&\int_{\spectrum} \rho(\xi-\la)\dm(\xi) \no\\
 &=&e^{-i(H_0-\la)T}\E\left(e^{-i(H_0-\la)d_l}\right),\
 l=1,2,\ldots ,\label{eq:FTdH}\\
 \hat g_0(\la-H_0)\Pc&=&\int_{\spectrum} \hat
g_0(\la-\xi)\dm(\xi) \no\\ &=&\int_0^T
g_0(t)e^{-i(\la-H_0)t}\Pc dt ,\label{eq:FTH0} \ea where
$\dm(\xi)$ is the absolutely continuous part of the spectral
measure of $H_0 . $

\begin{rmk} {\bf Conditions implying (H7')}\ If $H_0=-\Delta+V(x)$
 is a Schr\" odinger operator with
potential, $V(x)$, which decays sufficiently rapidly as $x$ tends
to infinity, then either \be\label{eq:H7suf1} \E\left(
e^{i\la_jd_l}\right)=0,\quad l=0,1,\ldots \ {\rm and}\
j=1,2\ldots, m\ee
 or
 \be\label{eq:H7suf}\hat g_0(\la_j)=0,\
j=1,\dots, m \ee imply {\bf (H7')}, provided the mean and variance
of the random variables $d_0,d_1,\ldots ,$ are  finite. Note that
(\ref{eq:H7suf1}) is equivalent to adding the threshold,
 $\la_0=0$, of the continuous spectrum to the set of eigenvalues
$\{\la_k\ :\ k=1,2,\ldots ,m\}$ for which {\bf (H6')} must hold.
Hypothesis (\ref{eq:H7suf}) means that the perturbation should not
induce a resonant coupling between the bound states and the
threshold generalized eigenfunction associated with
$\la_0=0$.
\end{rmk}

In analogy with the case of discrete spectrum, we write the
solution of (\ref{eq:perturbed}) in the form
$$
\phi(t,x)=\sum_{j=1}^m a_j(t)\psi_j(x)+\Pc\phi(t,x).
$$
Recall that the
expected power vector $P(t)$ is defined as the column vector
$$P(t)=\left(\E (\overline a_1 a_1(t)), \E (\overline a_2 a_2(t)),\ldots,\E (\overline a_m
a_m(t))\right) .$$ We denote by
$$P^{(n)}=P(t),\qquad t_{n-1}+T\le t <t_n$$ the expected power vector
after $n\ge 1$ defects (note that $P(t)$ is constant on the above
intervals).

We will show that the change in the power vector induced by each
defect can be expressed in terms of a power transmission matrix
\ba\label{eq:radtrans}
\cT_{\ve}\ &=&\ T_{disc,\ve}\  -\ \ve^2\diag[\gamma_1,\gamma_2,\ldots,\gamma_m]
 \nn\\
           &=&\ \I\ -\ \ve^2B\ -\ \ve^2\diag[\gamma_1,\gamma_2,\ldots,\gamma_m]
\ea
Recall that $T_{disc,\ve}=T_\ve=\I-\ve^2B$,
 displayed in (\ref{eq:nTdef}-\ref{eq:rBdef}), is
the power transmission matrix for systems governed by discrete
spectrum.
Each damping coefficient $\gamma_k>0,\ k=1,2,\ldots,m$ results
from the interaction between the corresponding bound state and the
radiation field. In contrast to the results in \cite{kw:mbs},
there are no contributions from bound state - bound state
interactions mediated by the continuous spectrum; these terms
cancel out by stochastic averaging.

\begin{rmk} For sufficiently small $\ve$ we have:
\be\label{eq:radtransnorm}\|T_{\ve}\|_1=1-\ve^2\min\{\gamma_1,\gamma_2,\ldots,\gamma_m\}<1\ee
\end{rmk}

The damping coefficients 
 are given by: \be\label{eq:radgammak}
\gamma_k=\lim_{\eta\searrow 0}\left\|\hat
g_0(H_0-\la_k)\sqrt{\I-|\rho(H_0-\la_k-i\eta)|^2}\left(\I-\rho(H_0-\la_k-i\eta)\right)^{-1}\Pc[\beta\psi_k]\right\|^2>0,\ee
for all $k=1,2,\ldots,m.$ Here the operators which are functions
of $H_0$ are defined via the spectral theorem and $\I$ is the
identity on ${\cal H} .$ 

%
%

The following theorem is a generalization of our previous result on the
effect of a single defect on the mode-power distribution, adapted to the
case where the Hamiltonian has both discrete and continuous spectrum:

\begin{theo}\label{th:transmission} Consider the Schr\"odinger equation
 \be i\D_t\phi\ =\ H_0\phi\ +\ g(t)\beta\phi,\nn\ee
where $g(t)$ is a random function, defined in terms of $g_0(t)$, given by
 (\ref{eq:gdef}). Assume that  hypotheses
{\bf (H1'-H7')} hold. Consider initial conditions for (\ref{eq:perturbed})
 such that $\wpl\Pc\phi_0\in {\cal H} .$ Then there exists an
$\ve_0>0$ such that whenever $|\ve |\le\ve_0$ the solution of
(\ref{eq:perturbed}) satisfy:
\be\label{eq:transmission}
P^{(n+1)}=\cT_{\ve}P^{(n)}+\cO\left(\ve^3\right)+\cO\left(\frac{\ve}{\lan
nT\ran^{r}}\right),\qquad n=0,1,2,\ldots ,\ee where the matrix
$\cT_{\ve}$ is given in (\ref{eq:radtrans}) and
$r=\min\{r_1,r_2-1\}>1.$
\end{theo}

By applying this theorem successively we get the change over $n\ge
1$ defects: \be\label{eq:rad1}
P^{(n)}=\cT^n_{\ve}P(0)+\sum_{k=0}^{n-1}\cT_{\ve}^k
 \left(\cO(\ve^3)+\cO\left(\frac{\ve}{\lan
(n-k)T\ran^{r}}\right)\right) .\ee
Using $\|T_{\ve}\|_1<1$ and
$$\sum_{n=1}^\infty\lan nT\ran^{-r}<\infty$$
we can conclude that the last correction term in (\ref{eq:rad1})
is of order $\cO(\ve) .$\footnote{one can actually show that $
\sum_{k=0}^{n-1}\cT_{\ve}^k\cO\left(\frac{\ve}{\lan
(n-k)T\ran^{r}}\right)=\cO\left(\min\left\{\ve,\frac{\lan
nT\ran^{-r}}{\ve\gamma}\right\}\right).$ However, as
$n\rightarrow\infty$ the other correction term dominates and the
result of Theorem \ref{th:radtrain} cannot be improved.} As for
the other correction term we have two ways in computing its size.
The first is based on $\|\cT_{\ve}^k\|_1<1, $ and gives
$$\sum_{k=0}^{n-1}\cT_{\ve}^k\cO (\ve^3)\ =\ \cO (n\ve^3) .$$
The second is based on $$\sum_{k=0}^{n-1}\|\cT_{\ve}^k\|_1\le
(1-\|\cT_{\ve}\|_1)^{-1}\le {1\over\gamma\ve^2},$$ where
$\gamma=\min\{\gamma_1,\gamma_2,\ldots,\gamma_m\} ,$ and gives
$$\sum_{k=0}^{n-1}\cT_{\ve}^k\cO (\ve^3)\ =\ \cO (\ve\gamma^{-1}) .$$

We have proved the following theorem:

\begin{theo}\label{th:radtrain} Under the assumptions of Theorem
\ref{th:transmission}, the expected power vector after $n$ defects,
$n=1,2,\ldots,$ satisfies:
$$P^{(n)}=\cT_{\ve}^nP(0)+\cO\left(\min(\ve\gamma^{-1},n\ve^3)\right)+\cO (\ve)
.$$
Here, $\cT_\ve$ is the diffusion/damping power transmission
 matrix given in (\ref{eq:radtrans}).
\end{theo}

Moreover, the argument we used in the proof of Theorem
\ref{th:fixedtime} now gives
\begin{theo}\label{th:radfixedtime} Under the assumptions of Theorem
\ref{th:transmission}, the expected power vector at a fixed time
$t, 0\le t <\infty$ satisfies:
\be\label{eq:radfixedtime}P(t)=\cT_{\ve}^nP(0)+\cO (\ve^{4/5})
.\ee Here, $n$ is the integer part of $t/(T+M),\ T$ is the common
time span of the defects and $M$ is the mean of the identically
distributed random variables $d_0,d_1,\ldots .$
\end{theo}
The nicer form of the correction term in (\ref{eq:radfixedtime})
compared to (\ref{eq:Pt}) is due to the fact that
$\min(t\ve^3,\ve/\gamma)$ is now dominated by $\cO(\ve^{4/5}).$

In analogy with Corollary \ref{rmk:resultinterpretfinite} we have,
in the present context, the following limiting behavior:
\begin{cor} Under the assumption of theorem \ref{th:transmission} the following holds:
\be\label{eq:radlimit} \lim_{t\rightarrow
\infty}P(t)=\left\{\begin{array}{lr}
P(0),&{\rm if}\ t\ll\ve^{-2}\\
e^{-(B+\Gamma)\tau}P(0) & {\rm if}\ t=\tau\ve^{-2}\\
0,&{\rm if}\  t\gg \ve ^{-2},\ \ve\rightarrow 0
 \end{array}\right. ,
\ee where $B$ is displayed in (\ref{eq:rBdef}) and
$$\Gamma=\diag\left[\gamma_1,\gamma_2,\ldots,\gamma_m\right]>0$$
\end{cor}
\nit\un{Proof} Since $\cT_\ve=\mathbb{I}-\ve^2(B+\Gamma)$ and $B+\Gamma$ is self adjoint with
$$B+\Gamma\ge \min\{\gamma_k\ :\ k=1,2,\ldots m\}>0$$ we have
\be\label{eq:radcTlimit}
\lim_{n\rightarrow \infty}\cT_\ve^{(n)}=\left\{\begin{array}{lr}
\mathbb{I},&{\rm if}\ n\ll\ve^{-2}\\
e^{-(B+\Gamma)\tau} & {\rm if}\ n=\tau\ve^{-2}\\
0 &{\rm if}\ n\gg\ve^{-2}, \ve\rightarrow 0
 \end{array}\right. .
\ee This follows from writing $\cT_\ve$ in the basis which
diagonalizes $B+\Gamma$ and using the fact that all eigenvalues of
$B+\Gamma$ are strictly positive, see the proof of Corollary
\ref{rmk:resultinterpretfinite}.

Clearly, (\ref{eq:radcTlimit}) and Theorem \ref{th:radfixedtime}
imply the conclusion of the corollary.\ \ \ []

Note that on time scales of order $1/\ve^2$ the dynamical system is now equivalent to:
$$\D_\tau P(\tau) = (-B -\Gamma)P(\tau),$$
where $-B$ is a diffusion operator, see the discussion after
relation (\ref{eq:heat}), while $-\Gamma$ is a damping operator.

It remains to prove Theorem \ref{th:transmission}.

\nit\un{Proof of Theorem \ref{th:transmission}.} Consider one realization of
the random variables $d_0,d_1,\ldots .$ For this realization the system
(\ref{eq:perturbed}) is linear, Hamiltonian and
deterministic. It is well known that such systems have an unique solution,
$\phi(t)$, defined for all $t\ge 0$ and continuously differentiable with
respect to $t .$ Moreover
\be\label{eq:radsolbound}
\|\phi(t)\|\equiv \|\phi_0\|.
\ee
We decompose the solution in its
projections onto the bound states and continuous spectrum of the unperturbed
Hamiltonian:
\be
\phi(t,x) =  \sum_{j=1}^ma_j(t)\psi_j+\Pc\phi(t)
= \phi_b(t)+\phi_d(t),\label{eq:raddecomposition}
\ee
where $\phi_b$ and $\phi_d$ are, respectively, the bound and dispersive parts
of $\phi$:
\ba
\phi_b(t)&=&\sum_{j=1}^ma_j(t)\psi_j,\label{eq:radphib}\nn\\
\phi_d(t)&=& \Pc\phi(t)\label{eq:radphid}
\ea
 and
\be\label{eq:radort}\lan\phi_b(t),\phi_d(t)\ran\equiv 0.\ee
Note that (\ref{eq:radsolbound}) and (\ref{eq:radort}) imply
\be\label{eq:radphibdbound} \|\phi_b(t)\|\le \|\phi_0\|,
\qquad \|\phi_d(t)\|\le\|\phi_0\|,\ee
for all $t\ge 0$. Consequently,
\be\label{eq:radakbound}|a_k(t)|\le\|\phi_0\|,\ee
for all $t\ge 0.$

By inserting (\ref{eq:raddecomposition}) into (\ref{eq:perturbed}) and
projecting the later onto the bound states and continuous spectrum we get
the coupled system:
\ba
i \D_t a_k(t)&=&\lambda_k a_k(t)+\ve g(t)
\left\lan\psi_k, \beta\phi_b(t)\right\ran +
\ve g(t)\left\lan\psi_k, \beta\phi_d(t)\right\ran,
\label{eq:radak}\\
i \D_t \phi_d(t)\ &=&\ H_0 \phi_d(t)\ +\ \ve g(t)\Pc
\beta\phi_d(t)\ +\ \ve g(t)\Pc\beta\phi_b(t),\label{eq:radphiddif}
\ea where $k=1,2,\ldots,m .$ Duhamel's principle applied to (\ref{eq:radphiddif})
yields
\be\label{eq:radphidint}\phi_d(t)=e^{-iH_0t}\phi_d(0)-
i\ve\int_0^tg(s)e^{-iH_0(t-s)}\Pc\beta\phi_d(s)ds
-i\ve\int_0^tg(s)e^{-iH_0(t-s)}\Pc\beta\phi_b(s)ds .\ee In a manner analogous to
the one in \cite{kn:CoSo} we are going to isolate $\phi_d$ in
(\ref{eq:radphidint}). Consider
the following two operators acting on $C(\R^+,Domain(\wpl))$ respectively $C(\R^+,{\cal H})$, the
space of continuous functions on positive real numbers with values
in $Domain(\wpl)$ respectively ${\cal H}$: \ba
K^+[f](t)&=&\int_0^tg(s)\wmi e^{-iH_0(t-s)}\Pc\beta\wpl f(s) ds\label{eq:radKtpl}\\
K[f](t)&=&\int_0^tg(s)\wmi e^{-iH_0(t-s)}\Pc\beta
f(s)ds.\label{eq:radKt} \ea Then, by applying the $\wmi$ operator
on both sides of (\ref{eq:radphidint}) we get:
\be\label{eq:radphidmi}\wmi\phi_d(t)=\wmi e^{-iH_0t}\phi_d(0)-
i\ve K^+[\wmi\phi_d](t) -i\ve K[\phi_b](t) .\ee On $C(\R^+,{\cal
H})$ we introduce the family of norms depending on $\alpha\ge 0:$
\be\label{eq:rnorm} \|f\|_\alpha=\sup_{t\ge 0}\lan
t\ran^\alpha\|f(t)\| \ee and define the operator norm:
\be\label{eq:oprnorm} \|\cA\|_\alpha=\sup_{\|f\|_\alpha\le
1}\|\cA f\|_\alpha. \ee The local decay hypothesis {\bf (H3')}
together with {\bf (H4')} and {\bf (H5')} imply:
\begin{lem}\label{lem:Kop}
If $0\le \alpha\le r_1$ then there exists a constant $C_\alpha$
such that \ba
\|K^+\|_\alpha &\le &C_\alpha\no\\
\|K\|_\alpha&\le &C_\alpha .\no \ea
\end{lem}

\nit\un{Proof of Lemma \ref{lem:Kop}.} Fix $\alpha,\ 0\le
\alpha\le r_1$ and $f\in C(\R_+,Domain(\wpl))$ such that
$\|f\|_\alpha\le 1$. Then \ba \lan t\ran^\alpha \|K^+[f](t)\|& =
&\lan t\ran^\alpha \left|\int_0^t g(s)\wmi e^{-iH_0(t-s)}\Pc\beta\wpl f(s)ds\right|\nn\\
&\le &\lan t\ran^\alpha \int_0^t|g(s)| \|\wmi
e^{-iH_0(t-s)}\Pc\wmi\|\cdot\|\wpl\beta\wpl\|\cdot \|
f(s)\|ds\nn\\
&\le &\lan t\ran^\alpha {\cal C}\|\wpl\beta\wpl\|
\int_0^t\frac{|g(s)|}{\lan t-s\ran^{r_1}}\|f(s)\|ds ,\nn\ea
 where we used (H3'). Furthermore, from
 $\|f\|_\alpha\le 1$ and $\|\wpl\beta\wpl\|$
 bounded, we have
\ba \lan t\ran^\alpha \|K^+[f](t)\|&\le &C\lan
t\ran^\alpha\int_0^t\frac{|g(s)|}{\lan t-s\ran^{r_1}\lan
s\ran^\alpha}\lan s\ran^\alpha\|f(s)\|ds\nn\\ &\le &C\lan
t\ran^\alpha\|f\|_\alpha\int_0^t\frac{|g(s)|}{\lan
t-s\ran^{r_1}\lan s\ran^\alpha}ds\nn\\
&\le &C\lan t\ran^\alpha\sum_{\{j:
t_j<t\}}\int_{t_j}^{\min(t,t_j+T)}\frac{|g(s)|}{\lan
t-s\ran^{r_1}\lan s\ran^\alpha}ds.\nn\ea By the mean value theorem
$$\int_{t_j}^{\min(t,t_j+T)}\frac{|g(s)|}{\lan t-s\ran^{r_1}\lan
s\ran^\alpha}ds=\lan{t-\tilde t_j}\ran^{-r_1}\lan \tilde
t_j\ran^{-\alpha}\|g_0\|_1,$$ for some \be\label{eq:tildetj}
t_j\le \tilde t_j\le \min(t,t_j+T).\ee Hence \be \lan t\ran^\alpha
\|K^+[f](t)\| \le C\lan t\ran^\alpha\sum_{\{j: \tilde
t_j<t\}}\lan{t-\tilde t_j}\ran^{-r_1}\lan \tilde t_j\ran^{-\alpha}
 \label{eq:K+estimate} \ee
We claim that \be\label{eq:csumest}\sum_{\{j: \tilde
t_j<t\}}\lan{t-\tilde t_j}\ran^{-r_1}\lan \tilde
t_j\ran^{-\alpha}\le D_\alpha\lan t\ran^{-\alpha}\ee for some
constant $D_\alpha $ independent of $t.$ This is a consequence of
the fact that we are computing the convolution of two power-like
sequences. For a more detailed proof we decompose the sum into
two, first running for $\tilde t_j\le t/2$ and the second for
$t/2<\tilde t_j\le t. $ For the former we have : \ba \sum_{\{j:
\tilde t_j<t/2\}}\lan{t-\tilde t_j}\ran^{-r_1}\lan \tilde
t_j\ran^{-\alpha}&\le& \left\lan {t\over
2}\right\ran^{-r_1}\sum_{\{j: \tilde t_j<t/2\}}\lan \tilde
t_j\ran^{-\alpha}\nn\\
&\le& \left\lan {t\over 2}\right\ran^{-r_1}\sum_{\{j:
jT<t/2\}}\lan jT\ran^{-\alpha}\label{eq:firstsum}\\
&\le& \left\lan {t\over 2}\right\ran^{-r_1}D_\alpha\left\lan
{t\over 2}\right\ran^{\max(0,1-\alpha)}\le D_\alpha\lan
t\ran^{-\alpha},\nn\ea since $r_1>\max (1,\alpha)$ and  $\tilde
t_j\ge t_j\ge (j-1)T$, see (H3'), the hypotheses of this lemma,
respectively (\ref{eq:tildetj}) and (\ref{eq:tndef}). The
remaining part of the sum is treated similarly: \ba \sum_{\{j:
t/2<\tilde t_j\le t\}}\lan{t-\tilde t_j}\ran^{-r_1}\lan \tilde
t_j\ran^{-\alpha}&\le& \left\lan {t\over
2}\right\ran^{-\alpha}\sum_{\{j: t/2<\tilde t_j\le t/2\}}\lan
t-\tilde t_j\ran^{-r_1}\nn\\
&\le& \left\lan {t\over 2}\right\ran^{-\alpha}\sum_{\{k:
kT<t/2\}}\lan kT\ran^{-r_1}\label{eq:secondsum}\\
&\le& \left\lan {t\over 2}\right\ran^{-\alpha}D\le D_\alpha\lan
t\ran^{-\alpha},\nn\ea since $r_1>1$ and $t-\tilde t_j\ge kT$
where $k$ is such that $t_{k+j}=\max\{t_p:\ t_p\le t\},$ see
(\ref{eq:tildetj}) and (\ref{eq:tndef}).

Now (\ref{eq:firstsum}) and (\ref{eq:secondsum}) imply
(\ref{eq:csumest}) which replaced in (\ref{eq:K+estimate}) proves
the required estimate for the $K^+$ operator. For the $K$ operator
the argument is completely analogous.\ \ \ []

We are going to use Lemma \ref{lem:Kop} for $\alpha=0$ and
$\alpha=r_1$. For $C_0$ and $C_{r_1}$ defined in the Lemma, let
$$C_K=\max\left\{C_0,C_{r_1}\right\}$$
Then, for $\ve$ such that $C_K\ve <1$, the inverse operator
$(I-i\ve K^+)^{-1}$ exists and it is bounded in the norms
(\ref{eq:oprnorm}) for $\alpha=0$ and $\alpha=r_1$. Then
(\ref{eq:radphidmi}) implies:
 \ba
\wmi\phi_d(t)&=&\left(I-i\ve K^+\right)^{-1}\left[\wmi
e^{-iH_0t}\phi_d(0)\right](t)-i\ve\left(I-i\ve K^+\right)^{-1}K[\phi_b](t)\no\\
&=&\cO\left(\lan t\ran^{-r_1}\|\wpl\phi_d(0)\|\right)-i\ve K[\phi_b]+\cO
\left(\ve^2\|K[\phi_b]\|\right).\label{eq:radwmiphidex} \ea
Thus we have expressed the dispersive part, $\phi_d(t)$ as a functional
of the bound state part, $\phi_b(t)$.
Substitution of
(\ref{eq:radwmiphidex}) into (\ref{eq:radak}) gives, for $k=1,2,\dots$:
\ba \D_t
a_k(t)&=&-i\lambda_k a_k(t)-i\ve g(t)\sum_{j=1}^ma_j(t)
\left\lan\psi_k, \beta\psi_j\right\ran\no\\
&-&
\ve^2 g(t)\left\lan\wpl\beta\psi_k, K[\phi_b](t)\right\ran\label{eq:radakex}\\
&+&\ve g(t)\left(\cO (\|\wpl\phi_d(0)\|\lan t\ran^{-r_1})+
 \cO\left(\ve^2\|K[\phi_b]\|\right)\right) \ \
k=1,2,\ldots,m.\no \ea
In particular (\ref{eq:radakex}) implies \be\label{eq:radaknl}
\overline a_k(t_n)=e^{i\la_k(t_n-t_l)}\overline
a_k(t_l)+\ve\sum_{p=l}^{n-1}e^{i\la_k(t_n-t_p)}D_p(d_0,d_1,\ldots,d_p),
\ee for all $k=0,1,\ldots m, n\ge 2$ and $l<n.$ Here each constant
$D_p$ depends on the realization of $d_0,d_1,\ldots d_p$ and
does not depend on the realization of any other random variable.
In addition all $D_p$ are uniformly bounded by a constant
depending only on $C_K$ above and the initial condition $\phi(0).$
Hence \be\label{eq:radakpq}
a_k(t_n)=e^{-i\la_k(t_n-t_l)}a_k(t_l)+\cO (\ve|n-l|) \ee for all
$n,l=0,1,2,\ldots , $ and $k=1,2,\ldots,m.$

We multiply both sides of (\ref{eq:radakex}) with $\overline a_k$,
then add the resulting equation to its complex conjugate. Then
we integrate from $t_n$ to $t_n+T$ and obtain for $k=1,2,\ldots, m$
\be \overline a_k a_k(t_n+T)-\overline a_k a_k(t_n)\ =
 \ R_1 + R_2 + R_3,\label{eq:atensab}
\ee
where
\ba
R_1\ &=&\ -i\ve\sum_{j=1}^m\lan\psi_k, \beta\psi_j\ran
\int_{t_n}^{t_n+T}g(t)\overline a_k(t) a_j(t)dt
+c.c.\label{eq:R1def}\\
R_2\ &=&\
-\ve^2 \lan\wpl\beta\psi_k,
\int_{t_n}^{t_n+T}g(t)\overline a_k(t)K[\phi_b](t)dt\ran +c.c.\label{eq:R2def}\\
R_3\ &=&\ \cO\left(\ve\lan t_n\ran^{-r_1}\right)+\cO\left(\ve^3\right).
\label{eq:R3def}\ea

If we neglect the $R_2$ and $R_3$ in (\ref{eq:atensab}) we
are left with $R_1$, which is precisely the expression associated with the
power transfer in systems with discrete spectrum;
see Section \ref{se:norad}. Moreover
$R_3$ has  norm asserted in (\ref{eq:transmission}). So, it
remains to show of $R_2$ that
 \ba \lefteqn{\E\left(\lan\wpl\beta\psi_k,
\int_{t_n}^{t_n+T}g(t)\overline
a_k(t)K[\phi_b](t)dt\ran+c.c.\right)}\nn\\
&=&\gamma_kP_k^n+\cO (\lan nT\ran^{-r}) +\cO
(\ve),\label{eq:radfinish} \ea where $\gamma_k$ is given by
(\ref{eq:radgammak}) and $r=\min\{r_1,r_2-1\}>1.$

We use integration by parts. Let
 \be\label{eq:radtildeK} \tilde
K[\phi_b](t)\equiv\int_{t_n+T}^tg(s)e^{-i\la_k(s-t_n)}K[\phi_b](s)ds,\qquad
t_n\le t\le t_n+T. \ee
and note that $K[\phi_b](t_n+T)=0$.
 Lemma \ref{lem:Kop} together with
\be
g(s)=g_0(s-t_n),\  t_n\le s\le t_n+T,
\label{eq:g0trans}
\ee
 imply the existence of
a constant $C$ with the property:
\be\label{eq:radtildeKbound}
\|\tilde K[\phi_b](t)\|\le C\|g_0\|_1^2=C, \ee
 uniformly in
$t_n\le t\le t_n+T .$
Define
\be\label{eq:radAkdef}
A_k(t)=a_k(t)e^{i\la_k(t-t_n)}, \ee
for $k=1,2,\ldots,m .$ Note that
\be\label{eq:radAkid} A_k(t_n)=a_k(t_n)\ee
From (\ref{eq:radakex}) we have
 \be\label{eq:radAkbound} |\D_t
A_k(s)|\le C\ |\ve |\ |g_0(s-t_n)| \ee
for some constant $C$
independent of $s$ and $t_n\le s\le t_n+T .$ Now
\ba
\int_{t_n}^{t_n+T}g(t)\overline
a_k(t)K[\phi_b](t)dt&=&\int_{t_n}^{t_n+T}\overline
A_k(t)\D_t\tilde K[\phi_b](t)dt\no\\
&=&-\overline a_k(t_n)\tilde K[\phi_b](t_n)-\int_{t_n}^{t_n+T}
\D_t \overline A_k(t)\tilde K[\phi_b](t)dt\label{eq:radalmostthere}\\
&=&\overline
a_k(t_n)\int_{t_n}^{t_n+T}g(t)e^{i\la_k(t-t_n)}K[\phi_b](t)dt +\cO
(\ve).\no \ea
To further rewrite (\ref{eq:radalmostthere}) we note that
 for $t_n\le t\le t_n+T$
\ba
K[\phi_b](t)&=&\sum_{j=1}^m\int_{t_n}^ta_j(s)g(s)\wmi
e^{-iH_0(t-s)}\Pc\beta\psi_jds\no\\
&+&\sum_{l=0}^{n-1}\sum_{j=1}^m\int_{t_l}^{t_l+T}a_j(s)g(s)\wmi e^{-iH_0(t-s)}\Pc\beta\psi_jds
\label{eq:radKphibex1}\ea
An integration by parts similar to the one above and use of (\ref{eq:g0trans})
 leads to:
\ba
\lefteqn{\int_{t_l}^{t_l+T}a_j(s)g(s)\wmi e^{-iH_0(t-s)}\Pc\beta\psi_jds=}\no\\
&=&a_j(t_l)\int_{t_l}^{t_l+T}g(s)e^{-i\la_j(s-t_l)}\wmi
e^{-iH_0(t-s)}\Pc\beta\psi_jds+\cO\left(\frac{\ve}{\lan
t-t_l-T\ran^{r_1}}\right)\no\\
&=&a_j(t_l)\wmi \hat
g_0(\la_j-H_0)e^{-iH_0(t-t_l)}\Pc\beta\psi_j+\cO\left(\frac{\ve}{\lan
t-t_l-T\ran^{r_1}}\right),\label{eq:radwhatever} \ea and \ba
\lefteqn{\int_{t_n}^ta_j(s)g(s)\wmi e^{-iH_0(t-s)}\Pc\beta\psi_jds}\no\\
&=&a_j(t_n)\int_{t_n}^{t}g(s)e^{-i\la_j(s-t_n)}\wmi
e^{-iH_0(t-s)}\Pc\beta\psi_jds+\cO
(\ve).\label{eq:radwhatever1}\ea By plugging
(\ref{eq:radwhatever}-\ref{eq:radwhatever1}) in
(\ref{eq:radKphibex1}) we get \ba
K[\phi_b](t)&=&\sum_{j=1}^ma_j(t_n)\int_{t_n}^tg(s)e^{-i\la_j(s-t_n)}\wmi
e^{-iH_0(t-s)}\Pc\beta\psi_jds\no\\
&+&\sum_{j=1}^m\sum_{l=0}^{n-1}a_j(t_l)\wmi \hat
g_0(\la_j-H_0)e^{-iH_0(t-t_l)}\Pc\beta\psi_j+\cO (\ve),
\label{eq:radKphibex}\ea where to estimate the error we used the
fact that the series $\sum_l\lan t-t_l-T\ran^{-r_1}$ is convergent
and uniformly bounded in $t$.

We now substitute  (\ref{eq:radKphibex}) into the right hand side of
  (\ref{eq:radalmostthere}) and obtain
\ba
\lefteqn{\int_{t_n}^{t_n+T}g(t)\overline a_k(t)K[\phi_b](t)dt=\cO (\ve)}\no\\
&+&\sum_{j=1}^m\overline
a_k(t_n) a_j(t_n)\int_{t_n}^{t_n+T}g(t)e^{i\la_k(t-t_n)}\int_{t_n}^tg(s)e^{-i\la_j(s-t_n)}\wmi
e^{-iH_0(t-s)}\Pc\beta\psi_jdsdt\no\\
&+&\sum_{j=1}^m\sum_{l=0}^{n-1}\overline a_k(t_n)a_j(t_l)\wmi \hat
g_0(H_0-\la_k)\hat g_0(\la_j-H_0)e^{-iH_0(t_n-t_l)}\Pc\beta\psi_j.
\label{eq:radaathere} \ea Based on (\ref{eq:radaknl}) we can
replace $\overline a_k(t_n)a_j(t_l)$ in (\ref{eq:radaathere}) with
\ba\overline a_k(t_n)a_j(t_l)&=&e^{i\la_k(t_n-t_l)}\overline
a_k(t_l)a_j(t_l)+error(l,j)\label{eq:backward}\\
error(l,j)&=&\ve\sum_{p=l}^{n-1}e^{i\la_k(t_n-t_p)}D_p\wmi \hat
g_0(H_0-\la_k)\hat
g_0(\la_j-H_0)e^{-iH_0(t_n-t_l)}\Pc\beta\psi_j.\nn\ea Taking into
account that $t_n-t_{n-1}=d_n+T$ and the fact that
$t_{n-1}-t_l,\ D_p,\ l\le p\le n-1$ do not depend on $d_n ,$
the expected value of the error can be rewritten as \ba
\lefteqn{\E(error(l,j))=}\nn\\&=&\ve\sum_{p=l}^{n-1}\E\left(\wmi
e^{i(\la_k-H_0)(t_n-t_{n-1})}e^{i\la_k(t_{n-1}-t_p)}D_p\hat
g_0(H_0-\la_k)\hat
g_0(\la_j-H_0)e^{-iH_0(t_{n-1}-t_l)}\Pc\beta\psi_j\right)\nn\\
&=&\ve\sum_{p=l}^{n-1}\wmi
\rho(H_0-\la_k)\E\left(e^{i\la_k(t_{n-1}-t_p)}D_p\hat
g_0(H_0-\la_k)\hat
g_0(\la_j-H_0)e^{-iH_0(t_{n-1}-t_l)}\Pc\beta\psi_j\right)\nn\\
&=&\ve\sum_{p=l}^{n-1}\E\left(e^{i\la_k(t_{n-1}-t_p)}D_p\wmi\rho(H_0-\la_k)\hat
g_0(H_0-\la_k)\hat
g_0(\la_j-H_0)e^{-iH_0(t_{n-1}-t_l)}\Pc\beta\psi_j\right).\label{eq:errorexp}\ea
By applying the ${\cal H}$ norm to (\ref{eq:errorexp}), commuting
the norm with both summation and expected value and using {\bf
(H7')} we get: \be\label{eq:errorest} \|\E(error(l,j))\|\le |\ve
|\frac{{\cal C}(n-l)}{\lan t_{n-1}-t_l\ran^{r_2} }\le C\ |\ve
|\lan(n-l)T\ran^{1-r_2}.\ee Since $r_2>2$ the summation over $l$
and $j$ of all the errors will have an $\cO (\ve)$ size. By this
argument (\ref{eq:radaathere}) becomes: \ba
\lefteqn{\E\left(\left\lan\wpl\beta\psi_k,
\int_{t_n}^{t_n+T}g(t)\overline
a_k(t)K[\phi_b](t)dt\right\ran+c.c.\right) =
 \sum_{j=1}^m\E(\overline
a_k(t_n) a_j(t_n))}\no\\
&\cdot&\left\lan\wpl\beta\psi_k,
\int_{t_n}^{t_n+T}g(t)e^{i\la_k(t-t_n)}\int_{t_n}^tg(s)e^{-i\la_j(s-t_n)}\wmi
e^{-iH_0(t-s)}\Pc\beta\psi_jds dt\right\ran+c.c.\no\\
&+&\sum_{j=1}^m\sum_{l=0}^{n-1}\E(\overline a_k(t_l)a_j(t_l))\no\\
&\cdot&\E\left( \left\lan\wpl\beta\psi_k,\wmi \hat
g_0(H_0-\la_k)\hat
g_0(\la_j-H_0)e^{i(\la_k-H_0)(t_n-t_l)}\Pc\beta\psi_j\right\ran\right)+c.c.\no\\
&+&\cO (\ve).\label{eq:radaaathere}\ea
 But {\bf (H6')} and the technique used to prove (\ref{eq:ICnp1}) imply
$$\E(\overline a_k(t_l)a_j(t_l))=\left\{\begin{array}{lr}P_k^{(l)} & {\rm for}\
k=j,\\ 0 & {\rm for}\ k\neq j\end{array}\right.$$ Moreover, an
argument similar to the one we used in
(\ref{eq:backward}-\ref{eq:errorest}) allows us to replace
$P^{(l)}$ by $P^{(n)}$ in (\ref{eq:radaaathere}) and incur an $\cO
(\ve)$ total error. Then, (\ref{eq:radaaathere}) becomes \ba
\lefteqn{\E\left(\left\lan\wpl\beta\psi_k,
\int_{t_n}^{t_n+T}g(t)\overline
a_k(t)K[\phi_b](t)dt\right\ran+c.c.\right)
=}\no\\&=&P_k^{(n)}\left\lan\wpl\beta\psi_k,\wmi\hat
g_0(H_0-\la_k)\hat g_0(\la_k-H_0)\Pc\beta\psi_k\right\ran\no\\
&+&P_k^{(n)} \left\lan\wpl\beta\psi_k,\wmi \hat g_0(H_0-\la_k)\hat
g_0(\la_k-H_0)
\E\left(\sum_{l=0}^{n-1}e^{i(\la_k-H_0)(t_n-t_l)}\Pc\right)\beta\psi_k
\right\ran+c.c.
\no\\
&+&\cO (\ve)\label{eq:radaaaathere}. \ea

We claim that\ba
\gamma_k^n&\stackrel{def}{=}&\left\lan\wpl\beta\psi_k,\wmi\hat
g_0(H_0-\la_k)\hat g_0(\la_k-H_0)\Pc\beta\psi_k\right\ran\no\\
&+&\left\lan\wpl\beta\psi_k,\wmi \hat g_0(H_0-\la_k)\hat
g_0(\la_k-H_0)
\E\left(\sum_{l=0}^{n-1}e^{i(\la_k-H_0)(t_n-t_l)}\Pc\right)\beta\psi_k
\right\ran+c.c.\no\\
&=&\gamma_k+\cO(\lan nT\ran^{1-r_2})\label{eq:gammakn}\ea where
$\gamma_k$ is given in (\ref{eq:radgammak}). (\ref{eq:gammakn})
replaced in (\ref{eq:radaaaathere}) gives (\ref{eq:radfinish})
which finishes the proof of this Theorem.

To prove (\ref{eq:gammakn}) we first find a simpler expression for
the expected value operator involved. Since $\{d_j\}_{j\ge0}$ are
independent, identically distributed with common characteristic
function, $\rho(\xi),$ using the definition of $t_n,\ n\ge 0,$ see
(\ref{eq:tndef}) and the spectral resolution of the operator
$H_0,$ see (\ref{eq:FTdH}), we have:
 \ba
\E\left(e^{i(\la_k-H_0)(t_n-t_l)}\Pc\right)&=&
\int_{\spectrum}\E (e^{i(\la_k-\xi)(t_n-t_l)})\ \dm(\xi)\no\\
&=&\int_{\spectrum}
\E(e^{i(\la_k-\xi)\sum_{j=l}^{n-1}(d_j+T)})\ \dm(\xi)\no\\
&=&\int_{\spectrum}\prod_{j=l}^{n-1}\E
(e^{i(\la_k-\xi)(d_j+T)})\ \dm(\xi)\no\\
&=&\int_{\spectrum}\rho^{n-l}(\xi-\la_k)\
dm(\xi)=\rho^{n-l}(H_0-\la_k)\Pc.
 \label{eq:radaaaaathere}
\ea Hence \ba\lefteqn{\wmi \hat g_0(H_0-\la_k)\hat g_0(\la_k-H_0)
\E\left(\sum_{l=0}^{n-1}e^{i(\la_k-H_0)(t_n-t_l)}\Pc\right)\beta}\no\\
&=&\wmi \hat g_0(H_0-\la_k)\hat g_0(\la_k-H_0)
\sum_{j=1}^{n}\rho^j(H_0-\la_k)\Pc\beta.\label{eq:rhoseries}\ea
But each operator term in (\ref{eq:rhoseries}) has its ${\cal
H}-$norm dominated by: \ba \lefteqn{\|\wmi \hat g_0(H_0-\la_k)\hat
g_0(\la_k-H_0)\rho^j(H_0-\la_k)\Pc\beta\|=}\no\\ &=&\|\wmi
\rho(H_0-\la_k)\hat g_0(H_0-\la_k)\hat g_0(\la_k-H_0)
\E(e^{-i(H_0-\la_k)(t_{j-1}-t_0)}\Pc\beta)\|\no\\
&\le &\frac{{\cal C}}{\lan
t_{j-1}-t_0\ran^{r_2}}\|\wpl\beta\|\le\lan
(j-1)T\ran^{-r_2}.\no\ea Now $r_2>2$ implies that the sequence
$1/\lan jT\ran^{r_2}$ is summable, and, by the dominant
convergence theorem, there exists: \ba \tilde
\gamma_k&=&\left\lan\wpl\beta\psi_k,\wmi\hat
g_0(H_0-\la_k)\hat g_0(\la_k-H_0)\Pc\beta\psi_k\right\ran\no\\
&+&\sum_{j=1}^\infty\left\lan\wpl\beta\psi_k,\wmi \hat
g_0(H_0-\la_k)\hat g_0(\la_k-H_0) \rho^j(H_0-\la_k)\Pc\beta\psi_k
\right\ran+c.c.\no\\ &=&\lim_{n\rightarrow\infty}\gamma_k^n.\no\ea
Moreover \ba
|\tilde\gamma_k-\gamma_k^n|&=&\sum_{j=n+1}^\infty\left\lan\wpl\beta\psi_k,\wmi
\hat g_0(H_0-\la_k)\hat g_0(\la_k-H_0)
\rho^j(H_0-\la_k\Pc\beta\psi_k \right\ran+c.c.\no\\ &\le &
2C\sum_{j=n}^\infty\lan jT\ran^{-r_2}\le D\lan
nT\ran^{1-r_2}.\label{eq:radgammacorr}\ea

Consider now, for $\eta >0,$ \ba\gamma_k^\eta
&=&\left\lan\wpl\beta\psi_k,\wmi\hat
g_0(H_0-\la_k)\hat g_0(\la_k-H_0)\Pc\beta\psi_k\right\ran\no\\
&+&\sum_{j=1}^\infty\left\lan\wpl\beta\psi_k,\wmi \hat
g_0(H_0-\la_k)\hat g_0(\la_k-H_0)
\rho^j(H_0-\la_k-i\eta)\Pc\beta\psi_k
\right\ran+c.c.\label{eq:gammaeta}\ea On one hand
\be\label{eq:rhoetaj}\rho^j(H_0-\la_k-i\eta)\Pc=
\E(e^{-\eta(t_j-t_0)}e^{-i(H_0-\la_k)(t_j-t_0)}\Pc)\ee and, by the
dominant convergence theorem, for all $j\ge 1$
$$\lim_{\eta\searrow
0}\rho^j(H_0-\la_k-i\eta)\Pc=\rho^j(H_0-\la_k)\Pc .$$ On the other
hand the series (\ref{eq:gammaeta}) is dominated uniformly in
$\eta$ by a summable series, because: \ba \lefteqn{\|\wmi \hat
g_0(H_0-\la_k)\hat g_0(\la_k-H_0)
\rho^j(H_0-\la_k-i\eta)\Pc\beta\|}\no\\ &=&\left\|
\int_0^T\int_0^Tdudsg_0(s+u)g_0(u)\E\left(e^{-\eta(t_j-t_0)}\wmi
e^{-i(H_0-\la_k)(t_j-t_0-s)}\Pc\beta\right)\right\|\no\\
&\le&\frac{{\cal C}e^{-\eta jT}}{\lan
t_j-t_o-T\ran^{r_1}}\|g_0\|_1\|\wpl\beta\|\le\lan
(j-1)T\ran^{-r_1}.\no\ea Here we used {\bf (H3')}, $\|g_0\|_1=1$
and $\|\wpl\beta\|$ bounded. Therefore, by the Weierstrass
criterion: \be\label{eq:gammaetalim}\lim_{\eta\searrow
0}\gamma_k^n=\tilde \gamma_k\ee

In addition (\ref{eq:rhoetaj}) implies
\ba\|\rho(H_0-\la_k-i\eta)\Pc\|&\le &
\E\left(e^{-\eta(t_1-t_0)}\|e^{-i(H_0-\la_k)(t_1-t_0)}\Pc\|\right)\no\\
&\le &e^{-\eta T}<1.\no\ea This makes
$(\mathbb{I}-\rho(H_0-\la_k-i\eta))\Pc$ invertible and given by
the Neumann series:
\be\label{eq:rhoinverse}(\mathbb{I}-\rho(H_0-\la_k-i\eta))^{-1}\Pc
=\sum_{j=0}^\infty\rho^j(H_0-\la_k-i\eta)\Pc.\ee Plugging
(\ref{eq:rhoinverse}) in (\ref{eq:gammaeta}) we have
\ba\gamma_k^\eta &=&\left\lan\beta\psi_k,\hat
g_0(H_0-\la_k)\hat g_0(\la_k-H_0)\Pc\beta\psi_k\right\ran\no\\
&+&\left\lan\beta\psi_k,\hat g_0(H_0-\la_k)\hat g_0(\la_k-H_0)
\rho(H_0-\la_k-i\eta)(\mathbb{I}-\rho(H_0-\la_k-i\eta))^{-1}\Pc\beta\psi_k
\right\ran+c.c.\no\ea A simple inner product manipulation shows
that: $$\gamma_k^\eta=\left\|\hat
g_0(H_0-\la_k)\sqrt{\I-|\rho(H_0-\la_k-i\eta)|^2}\left(\I-\rho(H_0-\la_k-i\eta)\right)^{-1}
\Pc[\beta\psi_k]\right\|^2.$$ Hence \be\label{eq:gammakexp} \tilde
\gamma_k=\lim_{\eta\searrow 0}\gamma_k^\eta=\gamma_k,\ee see also
(\ref{eq:gammaetalim}) and (\ref{eq:radgammak}).

Finally, (\ref{eq:gammakexp}) and (\ref{eq:radgammacorr}) give the
claim (\ref{eq:gammakn}). The theorem is now completely proven.\ \
\ []


\section{Comparison to stochastic approach}\label{se:comp}

In this section we want to compare our results with the {\it stochastic approach}
   in
\cite{kn:papa1,kn:papa2,kn:papa3,kn:kp}. We 
 view the results of this paper and those discussed in this section
  as complementary. 
 The results of this paper apply to the situation
when a known localized ``defect'', $g_0$, is  randomly distributed 
  in a manner which achieves averaged diffusive effect. The results of
  Papanicolaou {\it et. al.} apply to a  random medium, which is unknown
  and with assumptions about their distribution. 
  One of the key technical assumptions in this latter work
 is that the expected value of the
 randomness, {\it at any time},  is zero, {\it i.e.}\
  in our notation $\E ( g(t)) =0$.
In the results of this paper, we allow for $\E ( g(t))$ to vary with
 $t$. Indeed, for our train of
pulses (see (\ref{eq:gdef}) and figure \ref{fig:train}) $\E (
g(t))=0$ and implies $g_0(t)\equiv 0$, so unless we have the
  $g_0\equiv 0$, $\E (g(t))$ is generally different
from zero and time-dependent. 
 On the other hand, our hypothesis {\bf (H4)} has no
corresponding restriction in Papanicolaou {\it et. al.}'s theory.

Another important difference  is that our result applies on time
scales even larger than $1/\ve^2$, where $\ve$ is the size of the
randomness while the other results apply only on time scales up to
$1/\ve^2$. However, it appears that there is a striking similarity
between the two results on $1/\ve^2$ time scales. The train of
pulses we analyzed is closest to the stochastic process described in
\cite[Section 2]{kn:papa1} where both its values in the epochs
$[0,d_0+T],[d_0+T,d_0+T+d_1+T]\ldots ,$ and the epochs are now dependent on the
realizations of the {\em same} random variables, $d_0, d_1,\ldots .$ However,
if we assume that the radiation modes are not present, the dynamical system we 
investigate, (\ref{eq:radak}-\ref{eq:radphiddif}), 
is the one in \cite[Section 4]{kn:papa1}, see also \cite{kn:papa2}. These 
prevents us to use the formulas in the above papers. Nevertheless, we are 
going construct another stochastic perturbation, in the spirit of the one 
in \cite[Section 4]{kn:papa1}, for which we can compute the expected
power evolution using both theories. We find that the two results coincide
but keep in mind that while the example satisfies all our
hypothesis it does not satisfy one of theirs, see below.

\begin{figure}[h]
 \begin{center}
  \begin{picture}(340,90)(-10,0) \setlength{\unitlength}{1pt}
    \put(0,20){\line(1,0){30}}
      \multiput(0,20)(0,-2){10}{\line(0,-1){1}}
      \put(15,0){\vector(-1,0){15}}
      \put(15,0){\vector(1,0){15}}
      \put(0,5){\makebox(30,10){{\tiny $d_0$}}}
      \put(0,18){\line(0,1){4}}
    \put(30,0){\pulse}
       \put(30,50){\makebox(60,10){{\tiny $g_0(t-d_0)$}}}
    \put(90,20){\line(1,0){10}}
    \put(100,20){\line(1,0){20}}
      \multiput(100,20)(0,-2){10}{\line(0,-1){1}}
      \put(110,0){\vector(-1,0){10}}
      \put(110,0){\vector(1,0){10}}
      \put(100,5){\makebox(20,10){{\tiny $d_1$}}}
      \put(100,18){\line(0,1){4}}
      \put(90,20){\makebox(20,10){{\tiny $T+d$}}}
    \put(120,0){\pulse}
       \put(120,50){\makebox(60,10){{\tiny $g_0(t-(T+d)-d_1)$}}}
    \put(180,20){\line(1,0){20}}
    \put(200,20){\line(1,0){40}}
      \multiput(200,20)(0,-2){10}{\line(0,-1){1}}
      \put(220,0){\vector(-1,0){20}}
      \put(220,0){\vector(1,0){20}}
      \put(200,5){\makebox(40,10){{\tiny $d_2$}}}
      \put(200,18){\line(0,1){4}}
      \put(190,20){\makebox(20,10){{\tiny $2(T+d)$}}}
    \put(240,0){\pulse}
       \put(240,50){\makebox(60,10){{\tiny $g_0(t-2(T+d)-d_2)$}}}
    \put(300,18){\line(0,1){4}}
    \put(300,20){\makebox(25,10){{\tiny $3(T+d)$}}}
    \multiput(305,20)(5,0){3}{\circle*{1}}
    \put(320,20){\vector(1,0){10}}
      \put(320,10){\makebox(10,10){{\tiny $t$}}}

    \put(0,20){\vector(0,1){70}}
    \put(-10,20){\line(1,0){10}}
    \put(-10,20){\makebox(10,10){{\tiny $0$}}}
    \put(0,80){\makebox(20,10){{\tiny $g(t)$}}}
\end{picture}
\end{center}
\caption{Another train of short lived perturbations}
 \label{fig:train1}
\end{figure}
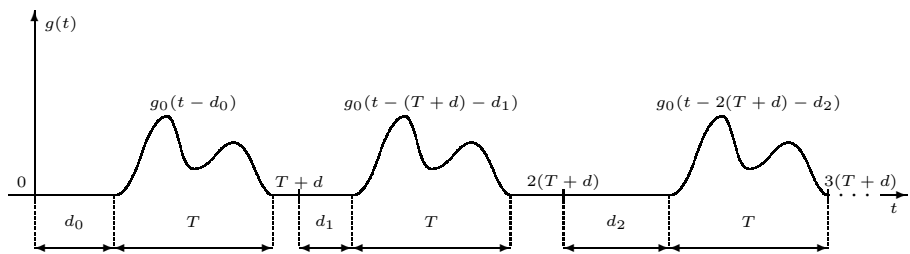

In addition to {\bf (H1)-(H4)}, suppose the random variables
$d_0,d_1,d_2,\ldots , $ can only take values in the interval
$[0,d]$ (this is clearly satisfied by the random variables constructed in the
previous section for finitely many modes) and denote by $\mu(t)$ the measure
induced by their distribution. Consider the positive real axis partitioned 
into ``epochs": $[0,T+d], [T+d,2(T+d)],\ldots $ of length $T+d$. 
 In each epoch a defect is placed at a distance $d_j$ from the starting
 point of the $j^{\rm th}$ epoch. Specifically,  
 the first defect is placed  at a distance
$d_0$ from $t=0$, the second at a distance $d_1$ from $t=T+d$,  
 $\ldots .$ Here $d_0,d_1,d_2,\ldots$ are realizations of the random
variables $d_0,d_1,d_2,\ldots $. That is, we will now consider equation
(\ref{eq:schrodinger}) with the perturbation given by:
\be\label{eq:comp2pulse}
g(t)=g_0(t-d_0)+g_0(t-(T+d)-d_1)+g_0(t-2(T+d)-d_2)+\ldots .
\ee
see figure \ref{fig:train1}.

Our result, Theorem \ref{th:train} applies without any modifications since
before each perturbation we have:
$$\E\left( a_k \overline a_j(l(T+d)+d_{l})\right)=\E\left( a_k \overline
a_j(l(T+d))\right)\E\left( e^{-i\Delta_{kj}d_{l}}\right)=0,$$ if
$k\neq j$. As for Theorem \ref{th:fixedtime} its proof is much
simplified and the error estimate improved because we now know how
many complete defects are going to appear up until the chosen time
$``t"$, namely $n=\lfloor t/(T+d)\rfloor$. The expected power at
time $t$ can differ from the one at time $n(T+d)$ by no more than
the size of the perturbation, $\ve$, since after each experiment
only a part or a single full defect can occur in between this time
slots. Hence: \be P(t)=P^(n)+\cO (\ve)=T_\ve^nP(0)+\cO
(n\ve^3)+\cO (\ve), \ee where the integer $n$ is such that
$n(T+d)\le t<(n+1)(T+d)$. To get closer to Papanicolaou {\it et.
al.}'s results, suppose
$$t=\tau /\ve^2,\qquad \tau\ge 0\ {\rm is\ fixed}$$
and pass to the limit $\ve\searrow 0$. We get \be\label{eq:ourr}
\lim_{\ve\searrow 0}P(t)=\lim_{\ve\searrow
0}T_\ve^{\left\lfloor\frac{\tau
}{(T+d)\ve^2}\right\rfloor}P(0)=\lim_{\ve\searrow
0}(\mathbb{I}-\ve^2B)^{\left\lfloor\frac{\tau
}{(T+d)\ve^2}\right\rfloor}P(0)=e^{\tau \tilde B}P(0) \ee where
$B$ is given in (\ref{eq:rBdef}) and \be\label{eq:tildeB} \tilde
B=-\frac{1}{(T+d)}B. \ee

Let us now apply Papanicolaou {\it et. al.} result to the above example. Note
that the manner in which 
 the perturbation is constructed makes the example very close to
that in \cite[Section 4]{kn:papa1}. But since the stochastic process is not
piecewise constant, one has to rely on more general form of their results such
as \cite[Remark 2 in Section 2]{kn:papa3}. The ODE system for the amplitude
vector, $a(t)=(a_1(t),a_2(t),\ldots )'$, is:
\ba
\D_ta(t)&=&Aa(t)-i\ve g(t)\alpha a(t),\label{eq:amplit}\\
a(0)&=&a(0),\no
\ea
where
\ba
A&=&-i\diag[\la_1,\la_2,\ldots],\no\\
\alpha&=&\left(\lan\psi_k,\beta\psi_j\ran\right)_{1\le k,j};\no
\ea
see also (\ref{eq:amplitudesystem}). This is a special case of system (2.27)
in \cite{kn:papa3} with
$$
\tilde M\equiv -ig(t)\alpha.
$$
Note that hypothesis (2.28) in \cite{kn:papa3} translates into
\be\label{eq:comp0mean}0=\E ( g(t)) =\int_0^dg_0(t'-s)d\mu(s),\quad
{\rm for\ all\ } t\ge 0\ee
where
$$t'=t-(T+d)\left\lfloor\frac{t}{T+d}\right\rfloor,$$
which generally implies the trivial case $g_0\equiv 0$. Hence for nontrivial
examples Papanicolaou {\it et. al.}'s theory is not rigorously applicable. We
are going to replace (\ref{eq:comp0mean}) with a milder one:

\be
\lim_{t\nearrow\infty}{1\over t}\int_0^t\E ( g(t))dt=0,\
\label{eq:ap-mean0} 
\ee
{\it i.e.}\ 
  the time average of the expected value of the perturbation is zero, 
 under which we can formally derive closed coupled power equations.
While the results of the stochastic approach do not apply to this example 
 because $\E ( g(t))\ne0$, our results do apply and it is reasonable to 
 conjecture
 that there is an extension of the stochastic approach to this case.
In the special case, where $g$ is given in (\ref{eq:comp2pulse}), the condition
 (\ref{eq:ap-mean0}) reduces 
 to $\hat g_0(0)=0$,

Let us compute their equation for the evolution of the powers, i.e. we prove
that their system of equations for the product of amplitudes:
$$\D_\tau \E ( a\otimes \overline a)= V\E (a\otimes\overline a) ,$$ where
$$\tau={t\over \ve^2}\ {\rm is\ fixed\ as}\ \ve\searrow 0.$$
gives a closed equation in powers, i.e.
\be\label{eq:diag}
V_{pq,pq'}=0,\qquad {\rm if}\ q\neq q'
\ee
and consequently for the powers $P(\tau)\equiv \diag \E ( a\otimes \overline a(\tau))$ we have
\be\label{eq:theirr}
\D_\tau P(\tau)=\tilde V P(\tau),
\ee
with
\be\label{eq:tildeV}
\tilde V_{pq}=V_{pq,pq}.
\ee
The main point is that $\tilde V$ coincides with $\tilde B$ in our result
(\ref{eq:ourr}); see also (\ref{eq:tildeB})and (\ref{eq:rBdef}).
Thus the two results agree on time scales of order $1/\ve^2$.

For the formula of $V$ we only have to replace $M$ in
\cite[equation (2.35)]{kn:papa3} by its complex conjugate
$\overline   M$ whenever it applies on the right part of the
tensor product, i.e. \ba V&=&\lim_{t\nearrow\infty}{1\over
t}\int_{t_0}^{t+t_0}\int_{t_0}^s\E\left( M(s)M(\sigma)\otimes
I+M(s)\otimes \overline M(\sigma)\right)d\sigma
ds\no\\&+&\lim_{t\nearrow\infty}{1\over
t}\int_{t_0}^{t+t_0}\int_{t_0}^s\E\left(M(\sigma)\otimes\overline
M(s)+I\otimes\overline M(s)\overline
M(\sigma)\right)d\sigma ds,\label{eq:V}\\
M_{pq}&=&-ie^{i\Delta_{pq}t}g(t)\alpha_{pq};\no \ea see also
\cite[relation (2.32)]{kn:papa3}. It will be clear from the
argument below that the limit in (\ref{eq:V}) does not depend on
$t_0$ (note that this in in fact a requirement for the validity of
the theory) so we are going to work with $t_0=0$. Although the
computation of $V$ has been done in \cite{kn:papa2} (denoted there
by $\overline V$) and then summarized in \cite[Section 3]{kn:kp}
we are not able to use them because they relied on the
stationarity of the process, see \cite[relation (2.2)]{kn:papa2}
which is not satisfied by our example. Nevertheless we have
component wise: \ba
V_{pq,pq'}&=&-\delta_{pq'}\sum_{r}\alpha_{pr}\alpha_{rq}\lim_{t\nearrow\infty}{1\over t}\int_0^t\int_0^s
e^{i\Delta_{pr}s}e^{i\Delta_{rq}\sigma}\E ( g(s)g(\sigma)) d\sigma ds\no\\
&+&\alpha_{pq}\alpha_{q'p}\lim_{t\nearrow\infty}{1\over t}\int_0^t\int_0^s
e^{i\Delta_{pq}s}e^{i\Delta_{q'p}\sigma}\E ( g(s)g(\sigma)) d\sigma ds\no\\
&+&\alpha_{pq'}\alpha_{qp}\lim_{t\nearrow\infty}{1\over t}\int_0^t\int_0^s
e^{i\Delta_{qp}s}e^{i\Delta_{pq'}\sigma}\E ( g(s)g(\sigma)) d\sigma ds\label{eq:Vpqpq'}\\
&-&\delta_{pq}\sum_{r}\alpha_{rp}\alpha_{q'r}\lim_{t\nearrow\infty}{1\over t}\int_0^t\int_0^s
e^{i\Delta_{rp}s}e^{i\Delta_{q'r}\sigma}\E ( g(s)g(\sigma)) d\sigma ds,\no
\ea
where we have used $\overline\alpha_{kj}=\alpha_{jk}$ due to the self-adjointness
of $\beta$ in $\alpha_{kj}=\lan\psi_k,\beta\psi_j\ran$ and the fact that $g(t)$
is real valued. Thus it is sufficient to compute $$
\int_0^t\int_0^s e^{i\Delta_{kj}s}e^{i\Delta_{jl}\sigma}\E ( g(s)g(\sigma)) d\sigma ds .
$$
Let us fix $t$ and suppose for the moment that $t=n(T+d)$. Then
\ba \lefteqn{\int_0^t\int_0^s
e^{i\Delta_{kj}s}e^{i\Delta_{jl}\sigma}\E ( g(s)g(\sigma)) d\sigma
ds=}\no\\&&\sum_{m=0}^{n-1}\int_{m(T+d)}^{(m+1)(T+d)}\int_0^s
e^{i\Delta_{kj}s}e^{i\Delta_{jl}\sigma}\E ( g(s)g(\sigma)) d\sigma ds \no\\
&=&\sum_{m=0}^{n-1}\int_{m(T+d)}^{(m+1)(T+d)}\int_{m(T+d)}^s
e^{i\Delta_{kj}s}e^{i\Delta_{jl}\sigma}\E ( g(s)g(\sigma)) d\sigma ds\label{eq:integral}\\
&+&\sum_{m=0}^{n-1}\int_{m(T+d)}^{(m+1)(T+d)} e^{i\Delta_{kj}s}\E ( g(s)) \int_0^{m(T+d)} e^{i\Delta_{jl}\sigma}\E ( g(\sigma)) d\sigma ds,\no
\ea
where we have used the fact that the random variable $g(s)$ and $g(\sigma)$ are
independent unless $s$ and $\sigma $ are in between the same epochs. Now
\ba
\int_0^{m(T+d)} e^{i\Delta_{jl}\sigma}\E ( g(\sigma)) d\sigma&=&\sum_{r=0}^{m-1}\int_{r(T+d)}^{(r+1)(T+d)}e^{i\Delta_{jl}\sigma}\E ( g_0(\sigma-r(T+d)-d_{r+1})) d\sigma\no\\
&=&\sum_{r=0}^{m-1}e^{i\Delta_{jl}r(T+d)}\int_0^{T+d}e^{i\Delta_{jl}\sigma}\int_0^d g_0(\sigma-s)d\mu(s) d\sigma\no\\
&=&\sum_{r=0}^{m-1}e^{i\Delta_{jl}r(T+d)}\int_0^de^{i\Delta_{jl}s}\int_{-s}^{T+d-s}e^{i\Delta_{jl}\sigma} g_0(\sigma )d\sigma d\mu(s)\no\\
&=&\sum_{r=0}^{m-1}e^{i\Delta_{jl}r(T+d)}\int_0^de^{i\Delta_{jl}s}d\mu(s)\int_{0}^{T}e^{i\Delta_{jl}\sigma} g_0(\sigma )d\sigma \no\\
&=&\sum_{r=0}^{m-1}e^{i\Delta_{jl}r(T+d)}\E ( e^{i\Delta_{jl}d_{r+1}})\hat g_0(-\Delta_{jl})\equiv 0,\no
\ea
where we used ${\rm supp}\ g_0\subset [0,T]$, $\E ( e^{i\Delta_{jl}d_{r+1}}) =0$ if $j\neq l$, see {\bf (H4)}, and the fact that $\hat g_0(0)=0$.

The only nonzero terms left in (\ref{eq:integral}) are of the form:
\ba
\lefteqn{\int_{m(T+d)}^{(m+1)(T+d)}\int_{m(T+d)}^s e^{i\Delta_{kj}s}e^{i\Delta_{jl}\sigma}\E ( g(s)g(\sigma)) d\sigma ds}\no\\
&=&e^{i\Delta_{kl}m(T+d)}\int_0^{T+d}\int_0^se^{i\Delta_{kj}s}e^{i\Delta_{jl}\sigma}\E ( g_0(s-d_m)g_0(\sigma-d_m)) d\sigma ds \no\\
&=&e^{i\Delta_{kl}m(T+d)}\int_0^{T+d}\int_\sigma ^{T+d}e^{i\Delta_{kj}s}e^{i\Delta_{jl}\sigma}\int_0^d g_0(s-\xi)g_0(\sigma-\xi)d\mu(\xi )ds d\sigma \no\\
&=&e^{i\Delta_{kl}m(T+d)}\int_0^d e^{i\Delta_{kl}\xi}\int{-\xi}^{T+d-\xi}e^{i\Delta_{jl}\sigma}g_0(\sigma)\int_{\sigma}^{T+d-\xi}e^{i\Delta_{kj}s}g_0(s)ds d\sigma d\mu(\xi) .\no
\ea
Now, the upper limit of the integrals with respect to $s$ and $\sigma$ can be
replaced by $\infty$ without modifying their values since
${\rm supp}\ g_0\subset [0,T]$ and $\xi\in [0,d]$. Hence they do not depend on
$\xi $ and by computing the integral with respect to the measure $d\mu(\xi)$
first we get:
$$\int_0^d e^{i\Delta_{kl}\xi}d\mu(\xi)=\E ( e^{i\Delta_{kl}d_m}) =\delta_{kl}$$
Knowing that $k=l$ in order to get a non zero result, we can now
compute the integrals with respect to $s$ and $\sigma$ using
(\ref{eq:2boundary}) without the complex conjugate part. In
conclusion we have \ba\lefteqn{\int_0^t\int_0^s
e^{i\Delta_{kj}s}e^{i\Delta_{jl}\sigma}\E ( g(s)g(\sigma)) d\sigma
ds =}\no\\&&\left\lfloor {t\over T+d}\right\rfloor
{\delta_{kl}\over 2}\left(|\hat
g_0(-\Delta_{kj})|^2+{i\over\pi}{\rm
P.V.}\int_{-\infty}^\infty\frac{|g(\mu)|^2}{\mu+\Delta_{kj}}d\mu\right)+\cO (1)
\no\ea where the correction is needed for $t\neq n(T+d)$.
Consequently \ba
\lefteqn{\lim_{t\nearrow\infty}\int_0^t\int_0^s e^{i\Delta_{kj}s}e^{i\Delta_{jl}\sigma}\E ( g(s)g(\sigma)) d\sigma ds}\no\\
&=& {\delta_{kl}\over 2(T+d)}\left(|\hat g_0(-\Delta_{kj})|^2+{i\over\pi}{\rm P.V.}\int_{-\infty}^\infty\frac{|g(\mu)|^2}{\mu+\Delta_{kj}}d\mu\right).\label{eq:fcomp}
\ea

Replacing (\ref{eq:fcomp}) in the formula (\ref{eq:Vpqpq'}) it is easy to see
that $V_{pq,pq'}=0$ unless $q=q'$ and in the later case the first and the
fourth terms in (\ref{eq:Vpqpq'}) are complex conjugate which is true for the
second and third terms also. Simple arithmetic leads to
$$\tilde V_{pq}\equiv V_{pq,pq}=\tilde B_{pq}$$
where $\tilde B$ is given by (\ref{eq:tildeB}) and (\ref{eq:rBdef}).

In conclusion, on time scales of order $1/\ve^2$ our results for
the example in this section coincides with the one obtained by
Papanicolaou {\it et. al.} in the series of papers
\cite{kn:papa1,kn:papa2,kn:papa3,kn:kp}. As mentioned earlier, although  
our result applies directly, the stochastic approach
  requires the $\E(g(t))=0$ for all $t$.

\section{Appendix: Properties of the power transmission matrix}\label{app:Bop}

In this section we prove the properties of the matrix (linear
operator) $B$ we used in Corollaries
\ref{rmk:resultinterpretfinite} and
\ref{rmk:resultinterpretinfinite}. Recall that $B$ is given by
(\ref{eq:rBdef}) and is irreducible, see the discussion before
Corollary \ref{rmk:resultinterpretinfinite}. We note that
(\ref{eq:rBdef}) implies in particular that:
\begin{enumerate}
\item all the components of $B$ are real;
\item $b_{ii}\ge 0$ for all $i=1,2,\ldots ;$
\item $b_{ij}\le 0$ for all $i,\ j,\ i\neq j;$
\item $\sum_jb_{ij}=0$ or equivalently $b_{ii}=-\sum_{j,j\neq
i}b_{ij}$ for all $i=1,2,\ldots .$
\end{enumerate}
\begin{lem}\label{lem:Bfinite}
If the dimension of $B$ is finite, say $m,$ then $B$ is a
nonnegative, self adjoint matrix having $0$ as a simple eigenvalue
with corresponding normalized eigenvector: $$
r_0={1\over\sqrt{m}}(1,1,\ldots ,1)'$$
\end{lem}

\nit\un{Proof.} Since all components of $B$ are real, self
adjointness is equivalent to $$b_{jk}=b_{kj},\qquad\forall j,\ k,\
j\neq k.$$ From (\ref{eq:alphadef}) we have
$$\left|\alpha_{jk}\right|^2=\left|\overline\alpha_{kj}\right|^2=\left|\alpha_{kj}\right|^2,$$
where $\overline\alpha $ denotes the complex conjugate of the
complex number $\alpha .$

Now because $\hat g_0$ is the Fourier transform of a real valued
function, see (\ref{eq:FT}) and (H3), and because
$\Delta_{kj}=-\Delta_{jk},$ see (\ref{eq:Deltadef}), we have
$$\left|\hat g_0(-\Delta_{jk})\right|^2=\left|\hat
g_0(\Delta_{kj})\right|^2=\left|\overline{\hat
g_0(-\Delta_{kj})}\right|^2=\left|\hat
g_0(-\Delta_{kj})\right|^2.$$ Hence, for all $j\neq k$
\be\label{eq:appsa}b_{jk}=-\left|\alpha_{jk}\right|^2\left|\hat
g_0(-\Delta_{jk})\right|^2=\left|\alpha_{kj}\right|^2\left|\hat
g_0(-\Delta_{kj})\right|^2=b_{kj},\ee rendering $B$ self
adjoint\footnote{Identity (\ref{eq:appsa}) does not rely upon $B$
having a finite dimension. Therefore it is valid even when $B$ has
infinite dimension.}.

In order to prove that $B$ is nonnegative, consider an arbitrary
vector $X=(X_1,X_2,\ldots ,X_m)'$ and let $X^*=(\overline
X_1,\overline X_2,\ldots ,\overline X_m)$ denote its adjoint. Then
\ba X^*BX&=&\sum_{i,j=1}^mb_{ij}\overline X_iX_j=
\sum_{i=1}^mb_{ii}|X_i|^2+\sum_{i,j,i\neq j}b_{ij}\overline X_iX_j\no\\
&=&-\sum_{i,j,i\neq j}b_{ij}|X_i|^2+\sum_{i,j,i\neq
j}b_{ij}\overline X_iX_j\label{eq:apppos}\\
&=&\sum_{i,j,i<j}|b_{ij}|\cdot |X_i-X_j|^2\ge 0,\no \ea where we
used properties 3. and 4. above. Hence $B$ is nonnegative.

Now, if $Y=Br_0$ then
$$Y_i={1\over\sqrt{m}}\sum_{i,j}^mb_{ij}=0,$$ by property 4.
Consequently $0$ is an eigenvalue for $B$ with corresponding
eigenvector $r_0.$

To prove that $0$ is a simple eigenvalue we use the irreducibility
of $B.$ On the set of components $\{1,2,\ldots ,m\}$ of vectors in
$\mathbb{C}^m$ we define the following relation:
\begin{defin}\label{def:coupled} We say that components $i$ and $i$ are always
coupled to zeroth order.

We say that components $i,j$ are coupled to first order if
$b_{ij}\neq 0.$

We say that components $i,j$ are coupled to $n^{\rm th},\ n\ge 2$
order if there exists a sequence of components $k_1,k_2,\ldots ,
k_{n-1},$ such that the pairs $1,k_1;\ k_1,k_2;\ldots ;k_{n-1},j;$
are all coupled to first order.

We say that components $i,j$ are coupled if they are coupled to
any order.
\end{defin}
It is easy to show that ``to be coupled" is an equivalence
relation on the set of components $\{1,2,\ldots ,m\}.$ Hence it
induces a partition of the components.\bigskip

\un{Claim 1.} If $B$ is irreducible the above partition is
trivial.\bigskip

Indeed, if we assume contrary the partition is formed by at least
two proper subsets of the set of components $\{1,2,\ldots ,m\}.$
By a reordering of the components, i.e. a reordering of the
standard basis vectors in $\mathbb{C}^m,$ we can assume assume
that the partition is formed by:
$$\{1,2,\ldots ,m_1\},\ \{m_1+1,m_1+2,\ldots ,m_2\},\ldots $$
Then $b_{ij}=0$ whenever $i,j$ fall in different subsets of the
partition, otherwise they would be coupled. Consequently, $B$ has
the form:
$$B=\diag\left[B_1,B_2,\ldots\right],$$
where $B_1$ is a $m_1\times m_1$ matrix, $B_2$ is a $m_2\times
m_2$ matrix, etc. But these contradict the irreducibility of $B,$
see also the discussion before Corollary
\ref{rmk:resultinterpretfinite}.\bigskip

\un{Claim 2.} If $X=(X_1,X_2,\ldots ,X_m)'$ is a zero eigenvector
for $B$ and $i,j$ are coupled then $X_i=X_j.$\bigskip

Indeed, $X^*BX=0$ because $BX=0,$ and (\ref{eq:apppos}) implies
\be\label{eq:appzero}\sum_{i,j,i<j}|b_{ij}|\cdot |X_i-X_j|^2=
0.\ee If $i,j$ are coupled to the first order then by definition
$b_{ij}\neq 0$ and we must have $X_i=X_j$ in order for
(\ref{eq:appzero}) to hold. By induction on the order of coupling
one obtains the result of the claim.

Finally, Claim 1 and the irreducibility of $B$ imply that all
components are coupled. Then Claim 2 implies that all components
of a zero eigenvector must be equal. Hence all zero eigenvectors
are parallel to $r_0.$ Since $B$ is self adjoint this means that
$0$ is a simple eigenvalue.\ \ \ []

\begin{lem}\label{lem:Binfinite1} If $B$ is infinite dimensional,
then $B$ is a bounded linear operator on $\ell^1$ with $\|B\|_1\le
2.$ In addition, for $|\ve |\le 1,$ the operator
$T_\ve=\mathbb{I}-\ve^2B$ transforms positive vectors (i.e.
vectors with all components positive) into positive vectors and
conserves their $\ell^1$ norm.\end{lem}

\nit\un{Proof.} It is well known that $B=(b_{ij})_{1\le
i,k<\infty}$ is a bounded linear operator on $\ell^1$ iff there
exists a constant $C\ge 0$ such that: \be\label{eq:appl1norm}
\sum_{i=1}^\infty |b_{ij}|\le C,\qquad\forall j=1,2,\ldots\ee In
this case$\|B\|_1\le C.$ We are going to show that for $B$ given
by (\ref{eq:rBdef}) we can choose $C=2$ in (\ref{eq:appl1norm}).

Indeed, let us fix an arbitrary $j\in\{1,2,\ldots\}$ and consider
the $j^{\rm th}$ vector in the standard basis of $\ell^1:$
\be\label{eq:appXdef} X=\left(X_1,X_2,\ldots\right)',\quad
X_i=\left\{\begin{array}{lr} 0&{\rm if}\ i\neq j\\ 1&{\rm if}\
i=j\end{array}\right.\ee Let \ba A&=&\left(a_{ij}\right)_{1\le
i,j<\infty}\no\\
a_{ij}&=&\alpha_{ij}\ \hat
g_0(-\Delta_{ij})=\lan\psi_i,\beta\psi_j\ran\int_{-\infty}^\infty
g_0(t)e^{i(\la_i-\la_j)t}dt.\label{eq:appAdef}\ea By a direct
calculation we have \ba \sum_{i=1}^\infty
|b_{ij}|&=&\sum_{i=1}^\infty\left|\overline X_i\sum_{k,p=1}^\infty
a_{ik}a_{kp}X_p-\overline{\sum_{k=1}^\infty
a_{ik}X_k}\sum_{k=1}^\infty a_{ik}X_k\right|\no\\
&=&\sum_{i=1}^\infty\left|\overline X_i(A\cdot
AX)_i-\left(\overline{AX}\right)_i\left(AX\right)_i\right|\label{eq:appl1est}\\
&\le &\sum_{i=1}^\infty |X_1|\cdot |(A\cdot
AX)_i|+\sum_{i=1}^\infty |(AX)_i|^2.\no\ea Clearly $X\in \ell^2,\
\|X\|_2=1.$ We are going to prove below that:\bigskip

\un{Claim 3.} $A$ is a bounded operator on $\ell^2$ with
$\|A\|_2\le 1.$\bigskip

Hence \be\label{eq:appAXest} \sum_{i=1}^\infty
|(AX)_i|^2=\|AX\|_2^2\le\|X\|_2^2=1,\ee while using
Cauchy-Buniakowski-Schwarz inequality we have: \ba
\sum_{i=1}^\infty |X_1|\cdot |(A\cdot AX)_i|&\le &
\left(\sum_{i=1}^\infty |X_i|^2\right)
^{1/2}\left(\sum_{i=1}^\infty |(A\cdot AX)_i|^2\right) ^{1/2}\no\\
&=&\|X\|_2\cdot\|A\cdot
AX\|_2\le\|A\|_2^2\|X\|_2\label{eq:appAAXest}\\
&\le &1.\no\ea By plugging in (\ref{eq:appAXest}) and
(\ref{eq:appAAXest}) into (\ref{eq:appl1est}) we
get\be\label{eq:appBcolest}\sum_{i=1}^\infty |b_{ij}|\le 2\ee and,
since $j$ was arbitrary, (\ref{eq:appl1norm}) holds with $C=2.$
Consequently, $B$ is a bounded linear operator on $\ell^1$ with
norm $\|B\|_1\le 2.$

Consider now $$T_\ve=\mathbb{I}-\ve^2B,\qquad
T_\ve=\left(t_{ij}\right)_{1\le i,j<\infty}.$$ Then for $i\neq j,$
$$t_{ij}=-\ve^2b_{ij}\ge 0.$$ Note that by (\ref{eq:appsa}) we
also have: \be\label{eq:appTsa} t_{ji}=t_{ij},\qquad \forall
i,j.\ee On the other hand
$$t_{ii}=1-\ve^2b_{ii}=1-\ve^2\sum_{j,j\neq i} |b_{ij}|,$$ where we
used properties 3. and 4. above. Moreover $$\sum_{j=1}^\infty
|b_{ij}|=|b_{ii}|+\sum_{j,j\neq i}|b_{ij}|=\left|-\sum_{j,j\neq
i}b_{ij}\right|+\sum_{j,j\neq i}|b_{ij}|=2\sum_{j,j\neq
i}|b_{ij}|.$$ Using now (\ref{eq:appBcolest}) and (\ref{eq:appsa})
we have \be\label{eq:apphrsB}\sum_{j,j\neq i}|b_{ij}|={1\over
2}\sum_{j=1}^\infty |b_{ij}|={1\over 2}\sum_{j=1}^\infty
|b_{ji}|\le 1.\ee  Hence
$$t_{ii}=1-\ve^2\sum_{j,j\neq i} |b_{ij}|\ge 1-\ve^2\ge 0,\quad
{\rm if}\ |\ve |\le 1.$$ We also have:
$$\sum_{j=1}^\infty t_{ij}=t_{ii}+\sum_{j,j\neq i} t_{ij}=1-\ve^2\sum_{j,j\neq i}
|b_{ij}|+\ve^2\sum_{j,j\neq i} |b_{ij}|=1,$$ and by
(\ref{eq:appTsa}) \be\label{eq:appsum1} \sum_{i=1}^\infty
t_{ij}=\sum_{i=1}^\infty t_{ji}=1.\ee

Now let $$X=\left(X_1,X_2,\ldots\right)'\in \ell^1,\qquad X_j>0\
\forall j=1,2,\ldots$$ Then
$$ \left( T_\ve X\right)_i=\sum_{j=1}^\infty t_{ij}X_j>0$$ since
all terms in the sum are nonnegative with at least one being
strictly positive. Moreover
$$\|T_\ve X\|_1=\sum_{i=1}^\infty \left|(T_\ve
X)_i\right|=\sum_{i=1}^\infty\sum_{j=1}^\infty
t_{ij}X_j=\sum_{j=1}^\infty X_j\sum_{i=1}^\infty
t_{ij}=\sum_{j=1}^\infty X_j=\|X\|_1,$$ where we exchanged the
order of summation because we are dealing with convergent series
with nonnegative terms and we also used (\ref{eq:appsum1}).

The Lemma is now finished provided we prove Claim 3. Let
$$X=\left(X_1,X_2,\ldots\right)'\in \ell^2,\qquad \|X\|_2=1$$ be
arbitrary and denote by
$$Y(t)=\sum_{j=1}^\infty e^{i\la_j t}X_j\psi_j,\quad Y(t)\in {\cal
H},\quad\|Y(t)\|\equiv 1.$$ Then \ba
|X^*AX|&=&\left|\sum_{j,k=1}^\infty a_{jk}\overline
X_jX_k\right|=\left|\int_{-\infty}^\infty g_0(t)\lan {\textstyle
\sum_{j=1}^\infty} e^{i\la_j t}X_j\psi_j,{\textstyle
\sum_{k=1}^\infty}
e^{i\la_k t}X_k\beta\psi_k\ran dt\right|\no\\
&=&\left|\int_{-\infty}^\infty g_0(t)\lan Y(t),\beta Y(t)\ran
dt\right|\le \int_{-\infty}^\infty |g_0(t)|\cdot |\lan Y(t),\beta
Y(t)\ran |dt\no\\
&\le & \int_{-\infty}^\infty |g_0(t)|\cdot \|\beta\|_{{\cal
H}}\|Y(t)\|^2 dt=\|\beta\|_{{\cal H}}\int_{-\infty}^\infty
|g_0(t)|dt=\|\beta\|_{{\cal H}}\|g_0\|_{L^1}=1\no\ea where, at the
very end, we used (H2) and (H3).\ \ \ []

\begin{lem}\label{lem:Binfinite2} If $B$ is infinite dimensional,
then $B$ is a bounded, linear, self adjoint, nonnegative operator
on $\ell^2$ with spectral radius less or equal to $2.$ Moreover,
$0$ is not an eigenvalue for $B.$\end{lem}

\nit\un{Proof.} Because of (\ref{eq:appsa}) $B$ is symmetric on
$\ell^2.$ Consider the 2-form induced by $B$ on $\ell^2$: \ba
X^*BX&=&\sum_{i,j=1}^\infty b_{ij}\overline
X_iX_j=\sum_{i=1}^\infty b_{ii}
|X_i|^2+\sum_{i=1}^\infty\sum_{j,j\neq i}^\infty
b_{ij}\overline X_iX_j\no\\
&=&-\sum_{i=1}^\infty\sum_{j,j\neq i}^\infty b_{ij}\overline
|X_i|^2+\sum_{i=1}^\infty\sum_{j,j\neq i}^\infty b_{ij}\overline
X_iX_j=\sum_{i=1}^\infty\sum_{j,j\neq i}^\infty
|b_{ij}|(|X_i|^2-\overline X_iX_j),\label{eq:app2form}\ea where we
used properties 3. and 4. above. Hence \ba
X^*BX&=&\sum_{i=1}^\infty\sum_{j,j\neq i}^\infty
|b_{ij}|\left(|X_i|^2+\frac{|X_i|^2+|X_j|^2}{2}\right)=2\sum_{i=1}^\infty\sum_{j,j\neq
i}^\infty |b_{ij}|\cdot |X_i|^2\no\\
&\le &2\sup_i\left({\textstyle \sum_{j,j\neq i}}
|b_{ij}|\right)\sum_{i=1}^\infty |X_i|^2\le 2\|X\|_2^2,\no\ea
where we used (\ref{eq:apphrsB}). So the 2-form induced by $B$ is
bounded. Since $B$ is symmetric this implies that $B$ is a self
adjoint, bounded operator on $\ell^2$ with $\|B\|_2\le 2.$
Therefore its spectral radius is less or equal to 2.

Moreover, (\ref{eq:app2form}) implies
$$X^*BX=\sum_{i<j}|b_{ij}|\cdot |X_i-X_j|^2\ge 0.$$ On one hand this
shows that $B$ is nonnegative. On the other hand, together with
obvious generalizations of Claims 1 and 2 in Lemma
\ref{lem:Bfinite} for the case of infinitely but countable
components, it shows that if a zero eigenvector for the
irreducible operator $B$ exists then the eigenvector should have
all components equal. However such a vector is not in $\ell^2$
unless it is trivial. Therefore $0$ is not an eigenvalue for $B.$\
\ \ []

\bibliographystyle{plain}
\bibliography{ref}
\end{document}